\documentclass[journal,12pt,onecolumn,draftclsnofoot,]{IEEEtran}

\usepackage{cite}
\usepackage{booktabs}
\usepackage{tabularx}
\usepackage{threeparttable}
\usepackage{graphicx}
\usepackage{float}
\usepackage{subfig}
\usepackage{amssymb}
\usepackage{amsmath}
\usepackage{mathtools}
\usepackage{cite}
\usepackage[table]{xcolor}
\usepackage{graphicx}
\usepackage{caption}
\usepackage[absolute]{textpos}
\begin{document}
		\begin{textblock}{18}(2,0.5)
		\centering
		\noindent S. Sun and M. Tao, ``Characteristics of channel eigenvalues and mutual coupling effects for holographic reconfigurable intelligent surfaces," \textit{Sensors}, vol. 22, no. 14, pp. 5297, Jul. 2022.
	\end{textblock}
\title{Characteristics of Channel Eigenvalues and Mutual Coupling
Effects for Holographic Reconfigurable Intelligent Surfaces}
	\author{Shu Sun, \textit{Member, IEEE}, and Meixia Tao, \textit{Fellow, IEEE}
		\thanks{The authors are with the Department of Electronic Engineering, Shanghai Jiao Tong University, Shanghai 200240, China (e-mail: \{shusun, mxtao\}@sjtu.edu.cn).}
	}
	
	\maketitle
	
	\begin{abstract}
As a prospective key technology for the next-generation wireless communications, \textcolor{black}{reconfigurable intelligent surfaces (RISs)} have gained tremendous research interest in both the academia and industry in recent years. Only limited knowledge, however, has been obtained about the channel eigenvalue characteristics and spatial degrees of freedom (DoF) of systems containing \textcolor{black}{RISs}, especially when mutual coupling (MC) is present between the array elements. In this paper, we focus on the small-scale spatial correlation and eigenvalue properties excluding and including MC effects, \textcolor{black}{for RISs with a quasi-continuous aperture (i.e., holographic RISs)}. Specifically, asymptotic behaviors of far-field and near-field eigenvalues of the spatial correlation matrix of \textcolor{black}{holographic RISs} without MC are first investigated, where the counter-intuitive observation of a lower DoF with more elements is explained by leveraging the power spectrum of the spatial correlation function. Second, a novel metric is proposed to quantify the inter-element correlation or coupling strength in \textcolor{black}{RISs and ordinary} antenna arrays. Furthermore, in-depth analysis is performed regarding the MC effects on array gain, effective spatial correlation, and eigenvalue architectures for a variety of element intervals when \textcolor{black}{a holographic RIS works in the radiation and reception mode}, respectively. The analysis and numerical results demonstrate that a considerable amount of the eigenvalues of the spatial correlation matrix correspond to evanescent waves that are promising for near-field communication and sensing. More importantly, \textcolor{black}{holographic RISs} can potentially reach an array gain conspicuously larger than conventional arrays by exploiting MC, and MC has discrepant impacts on the effective spatial correlation and eigenvalue structures at the transmitter and receiver.
\end{abstract}
\vspace*{19pt}
\section{Introduction}
\label{Introduction}
The sixth-generation (6G) communication networks is envisioned to embrace numerous new use cases and challenging requirements~\cite{Tataria21P}. Among~the emerging candidate physical-layer technologies for 6G, \textcolor{black}{reconfigurable intelligent surfaces (RISs)}, sometimes also named large intelligent surfaces~\cite{Hu18TSP} and \textcolor{black}{holographic multiple-input multiple-output (MIMO)}~\cite{Huang20WC}, shows \textcolor{black}{promising} foreground in capacity and coverage enhancement, reconfigurable environment construction, \textcolor{black}{intelligent sensing and control}, and~\textcolor{black}{holographic communications}~\cite{Hu18TSP,Basar19Access,Huang20WC,Liu21CST,Huang21arXivPIEEE,Wan21TC,Roman21EJC,Chi22TAC}. We utilize \textcolor{black}{\textit{holographic RIS}~\cite{Wan21TC,Ma21PIERS}} herein as an umbrella term for the two-dimensional (2D) architectures with an element spacing equal to or smaller than half a wavelength of the carrier frequency, which can be perceived as an extension of massive MIMO~\cite{Marzetta10TWC} with the ultimate form of (approximately) spatially-continuous electromagnetic (EM) aperture~\cite{Pizzo20JSAC}. \textcolor{black}{Holographic RISs} can be employed at the base station (BS), user equipment (UE), and/or interacting objects in the propagation medium, and~is likely to bring immense advantages in not only spectral efficiency, energy efficiency, and~system scalability inherited from massive MIMO~\cite{Larsson14CM,Yan20Access,Yan21IoTJ}, but~also the manipulation of EM waves via anomalous reflection, refraction, polarization transformation, and~so on~\cite{Huang20WC}, reminiscent of 
metasurfaces in the optical regime~\cite{Holloway12APM,Sun_JOSA}. In~order to unleash the full potentials of \textcolor{black}{holographic RISs}, it is necessary to understand its fundamental properties, such as channel eigenvalues and spatial degrees of freedom (DoF). 

\subsection{Related~Work}
The use of dense antenna arrays for wireless communications was in fact explored over two decades ago (e.g., \cite{Chiurtu01ITW,Gesbert02GC,Wei05TWC}), and has resurged recently as a promising 6G \mbox{technology~\cite{Hu18TSP,Pizzo20JSAC,Huang20WC}.} Due to the limited element spacing and 2D (as opposed to one-dimensional) structure of a dense array, the~spatial correlation between array elements is not always non-zero even under isotropic scattering~\cite{Pizzo20JSAC,Bjornson20WCL,Sun21RISModel}. Among~the early work involving antenna spatial correlation, the~authors in~\cite{Shiu00TC} have considered spatial correlation among multi-element antennas, and~derived upper and lower capacity bounds taking into account antenna correlation. Fading correlation has also been studied in~\cite{Chuah02TIT} to examine the capacity growth with respect to the number of antenna elements. The~impact of antenna correlation on capacity has been examined for diverse signal inputs and correlation architectures in~\cite{Tulino05TIT}. Nevertheless, the~antenna element spacing is of half-wavelength or larger in~\cite{Shiu00TC,Chuah02TIT,Tulino05TIT}. The~capacity of spatially dense multiple antenna systems has been pioneered in~\cite{Chiurtu01ITW}, which showed that the capacity of such a system approaches a finite limit. Spatially dense MIMO arrays have also been studied in~\cite{Gesbert02GC}, where the array-gain normalized capacity has been analyzed and the performance of small wavelength-like MIMO arrays has been shown to be similar to that of arrays with larger apertures. The~asymptotic capacity associated with antenna arrays of fixed length has been analyzed in~\cite{Wei05TWC} for uniform linear antenna arrays, revealing that the asymptotic mutual information converges almost surely as the number of antenna elements approaches infinity due to the convergence of the eigenvalues. The~aforementioned work, however, focused mainly on the capacity and did not explicitly consider the spatial DoF that the arrays can~offer. 

The spatial DoF for \textcolor{black}{holographic RISs} in line-of-sight (LoS) environments have been investigated in~\cite{Dardari20JSAC}, which has revealed that the DoF can be larger than one even in strong LoS channel conditions, favorable for spatial multiplexing. In~\cite{Hu18TSP,Pizzo20JSAC}, the~asymptotic spatial DoF for sufficiently dense and large \textcolor{black}{holographic RISs} have been derived from the perspectives of channel capacity and Fourier plane-wave series expansion, respectively. The~achievable DoF for more common cases with finite element spacing and aperture areas has been investigated in~\cite{Sun21RISModel}, where it has been discovered that the spatial DoF decreases as the number of elements grows which seems counter-intuitive, but~the underlying causes have not been~identified. 

Due to the close proximity of neighboring elements in \textcolor{black}{a holographic RIS}, mutual coupling (MC) naturally arises. Broadly speaking, MC refers to EM interaction among the array elements and~can occur because of three mechanisms: direct space coupling between array elements, indirect coupling caused by near-by scatterers, and~coupling through feed network~\cite{Mailloux18Book,Singh13IJAP}. In~this paper, we mainly refer to MC stemming from the first mechanism. MC between array elements can be characterized by a conventional multi-port circuit model, such as an impedance matrix, an~admittance matrix, or~a scattering matrix. Theoretically, the~type of matrix used to represent the array network is not important since matrix transforms can be applied to change the type of matrix representation. There is abundant early research work on MC in antenna arrays (e.g.,~\cite{Lechtreck68TAP,Janaswamy02AWPL,Wallace04TWC,Sadat04APSS,Singh13IJAP,Neil18TAP,Chen18Access,Malmstrom18TEC,Wolosinski20EuCAP} and references therein), which considered linear arrays only or did not fix the array aperture while varying the element spacing. Aiming at exploring the physical characteristics of the emerging \textcolor{black}{holographic RISs} or similar structures for 6G communications, the~authors in~\cite{Williams20ICCW} have proposed an LoS communication model incorporating MC in the form of mutual impedance, and~studied the associated optimal beamforming strategies. The~model has then been extended in~\cite{Williams21ICC} to account for superdirectivity and MC effects as well as near-field propagation. An~end-to-end MC-aware communication model based on mutual impedances has been propounded in~\cite{Gradoni21WCL}, which is also EM-compliant and unit cell aware. In~\cite{Gao21WCSP}, a~circuit-based MIMO channel model considering antenna MC, size-related antenna equivalent circuit, and~channel small-scale fading has been proposed, and~statistical properties including temporal autocorrelation function and spatial
cross-correlation function, along with the effects of MC and size-related
antenna equivalent circuit on them have been revealed. The~authors in~\cite{Han22arXiv} have derived a beamforming vector of superdirective arrays based on a coupling matrix-enabled method, and~proposed an approach to obtain the coupling matrix via spherical wave~expansion. 

\subsection{Contributions}
Despite the aforementioned extensive research work, study on small-scale spatial characteristics, the~associated MC effects, and~the mechanism behind some unique phenomena for \textcolor{black}{holographic RISs} are still in the infancy. In~this paper, therefore, we carry out thorough investigation on the aspects above. Specifically, the~small-scale spatial correlation, eigenvalue behavior, and~spatial DoF for \textcolor{black}{holographic RISs} excluding and including MC are explored. The~major novelty and contributions of this article lie in the following~aspects: 
\begin{itemize}
	\item First, leveraging the block-Toeplitz with Toeplitz block (BTTB) matrix theory, we relate the eigenvalues of the spatial correlation matrix of the \textcolor{black}{holographic RIS} to the power spectrum of the spatial correlation function, and~explain the counter-intuitive phenomenon of seemingly lower spatial DoF with growing numbers of elements in \textcolor{black}{a holographic RIS} observed in our prior work~\cite{Sun21RISModel}, which has not been addressed in the literature to our best knowledge. This analysis also helps with distinguishing the spatial DoF corresponding to the far field and near field of \textcolor{black}{a holographic RIS}. 
	\item Second, we incorporate MC into the array response and spatial correlation matrix of \textcolor{black}{the holographic RIS considering realistic element sizes}, and~demonstrate the potential of \textcolor{black}{holographic RISs} to reach an extraordinary array gain that is significantly higher than conventional antenna arrays with concrete examples. The~results indicate that, different from the common belief that MC is always deleterious and should be avoided or compensated for, MC can be beneficial in boosting the array gain of \textcolor{black}{holographic RISs} even without sophisticated manipulation of excitation coefficients for the \mbox{array elements. }
	\item Furthermore, in-depth analysis and comparisons are performed regarding the MC effects on spatial correlation and the corresponding eigenvalue distributions for \textcolor{black}{holographic RISs} working in the transmitting (Tx) and receiving (Rx) modes, respectively, and~with various element intervals as well as source and load impedance values. A~metric named \textit{inter-element correlation/coupling strength indicator (ICSI) } is proposed to measure the amount of inter-element correlation/coupling within an array. Results show that the effects of MC are quite discrepant for Tx and Rx arrays, and~are also dependent upon element spacing, source and load impedance, among~other factors, necessitating comprehensive design and implementation considerations. 
\end{itemize}

Isotropic scattering is considered in this paper, since it is a typical type of environment involved in an enormous amount of theoretical research work, and~also encountered by low-frequency bands (e.g., sub-1 GHz) which are still crucial even in 6G to guarantee wide coverage and high reliability~\cite{Jeon21CM}, and~the corresponding results can serve as theoretical upper bounds for non-isotropic scattering scenarios. The~contributions in this paper unveil some fundamental channel eigenvalue features and practical spatial DoF of \textcolor{black}{holographic RISs}, and~provide valuable hints on channel estimation and beamforming strategies for \textcolor{black}{holographic RISs}~\cite{Sun21RIS,Williams21ICC,Han22arXiv}. 

\subsection{Article Outline and~Notation}
The remainder of this paper is organized as follows: in~Section~\ref{sec:sysModel}, we describe the system model, and~formulate array responses and spatial correlation excluding and including MC \textcolor{black}{for holographic RISs}. Analyses of asymptotic eigenvalue distributions of the spatial correlation matrix without and with MC are provided in Sections~\ref{sec:EV_noMC} and~\ref{sec:EV_MC}, respectively. Conclusions are drawn in Section~\ref{sec:Conclusions}. 

The following notations will be utilized throughout the paper: $\textbf{A}$ 
 for matrix, $\textbf{a}$ for column vector, $[\textbf{A}]_{i,j}$ for the $(i,j)$th entry of $\textbf{A}$; $\textbf{A}^\text{T}$, $\textbf{A}^*$, and~$\textbf{A}^\text{H}$ for the transpose, conjugate, and~Hermitian of $\textbf{A}$, respectively; $\det(\textbf{A})$ for determinant of the square matrix $\textbf{A}$, $\text{tr}(\textbf{A})$ for the trace of $\textbf{A}$, $\textbf{I}_N$ for the $N\times N$ identity matrix, while $\lceil a\rceil$ and $\lfloor a\rfloor$ for the ceiling and floor of the scalar $a$, respectively. 

\section{System~Model}
\label{sec:sysModel}
\begin{figure}
	\centering
	\includegraphics[width=0.6\columnwidth]{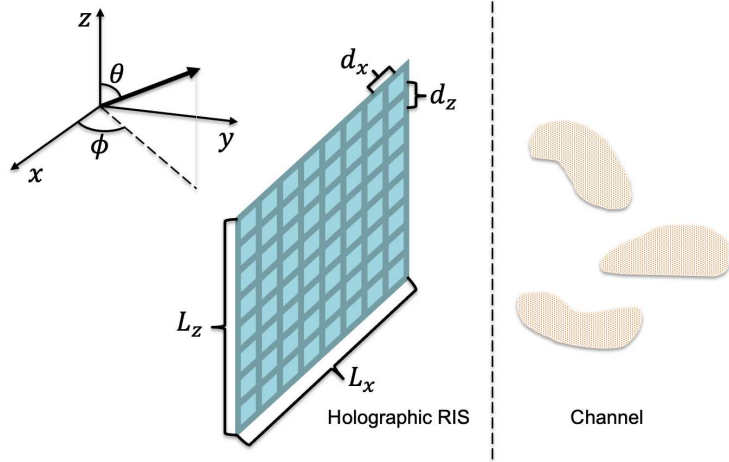}
	\caption{\textcolor{black}{System model and orientation of the holographic reconfigurable intelligent surface (RIS) with respect to the associated coordinate system. The~horizontal and vertical lengths of the holographic RIS are $L_x$ and $L_z$, with~element spacing of $d_x$ and $d_z$, respectively. The~azimuth and zenith angles are denoted by $\phi$ and $\theta$, respectively, and~the three irregular blocks in the channel represent random scatterers.}}
	\label{fig:SystemModel}
\end{figure}

We consider \textcolor{black}{a holographic RIS} in a wireless communication system where the \textcolor{black}{holographic RIS} can be employed at the BS, UE, and/or interacting objects that can radiate, receive, reflect, or~refract wireless signals, as~illustrated in Fig.~\ref{fig:SystemModel}. The~horizontal and vertical lengths of the \textcolor{black}{holographic RIS} are $L_x$ and $L_z$, with~element spacing of $d_x$ and $d_z$, respectively. \textcolor{black}{Each element in the holographic RIS is modeled as a cylindrical thin wire of perfectly conducting material, and~is connected to a tunable load, where the load can be a positive-intrinsic-negative diode whose inductance and capacitance are adaptable to reconfigure the response of each element~\cite{Gradoni21WCL}.}

The signal sent from or impinging on the \textcolor{black}{holographic RIS} is generally composed of a superposition of multipath components which can be regarded as a continuum of plane waves, hence the channel capturing small-scale fading can be expressed as~\cite{Sayeed02TSP}
\begin{equation}\label{eq:h1}
\textbf{h}=\int_{0}^{\pi}\int_{0}^{\pi}\tilde{s}(\phi,\theta)\textbf{a}(\phi,\theta)d\phi d\theta
\end{equation}

\noindent where $\tilde{s}(\phi,\theta)$ denotes the angular distribution function that contains the channel gain and phase shift corresponding to the direction $(\phi,\theta)$ with $\phi$ and $\theta$ representing the azimuth and zenith angles, respectively, while $\textbf{a}(\phi,\theta)$ is the array response vector. The~correlation function is given by
\begin{equation}\label{eq:R1}
\textbf{R}=\mathbb{E}\left\{\textbf{h}\textbf{h}^\text{H}\right\}=\int_{0}^{\pi}\int_{0}^{\pi}s(\phi,\theta)\textbf{a}(\phi,\theta)\textbf{a}^\text{H}(\phi,\theta)d\phi d\theta
\end{equation}

\noindent where $s(\phi,\theta)$ denotes the normalized spatial scattering function satisfying
\begin{equation}\label{eq:s1}
\mathbb{E}\left\{\tilde{s}(\phi,\theta)\tilde{s}^*(\phi^\prime,\theta^\prime)\right\}=s(\phi,\theta)\delta(\phi-\phi^\prime)\delta(\theta-\theta^\prime)
\end{equation}

\noindent and $\int_{0}^{\pi}\int_{0}^{\pi}s(\phi,\theta)d\phi d\theta=1$. Physically, the~presence of spatial correlation means that the signal strengths at different elements do not vary independently, but~may rise or fade~simultaneously. 

\subsection{Array Response and Spatial Correlation Excluding~MC}\label{sec:SC_noMC_theory}
For an array with $N$ elements, where each element has the same pattern function of $\textcolor{black}{p}(\phi,\theta)$ (Strictly speaking, if~MC exists, the~central elements and the ones near the array edges may not maintain the same element pattern when embedded in an array, even if their isolated element patterns are identical
~\cite{Mailloux18Book}. Nevertheless, it is possible to compensate for the pattern distortion via predetermined illumination, and~here we assume the same embedded element pattern for all the elements in an array. The~element pattern variation owing to MC is another topic and is deferred to future work), the~far-field radiation pattern of the array is expressed as
\begin{equation}\label{eq:f1}
f(\phi,\theta)=\sum_{n=1}^{N}w_n\textcolor{black}{p}(\phi,\theta)e^{j\kappa\bold{\hat{d}}\cdot\bold{d}_n}
\end{equation}

\noindent where $w_n$ denotes the complex excitation coefficient proportional to the current on the $n$-th element, $\kappa=2\pi/\lambda$ is the wavenumber with $\lambda$ being the carrier wavelength, $\bold{\hat{d}}$ represents the unit vector of the far-field direction $(\phi,\theta)$ in the spherical
coordinate system, and~$\bold{d}_n$ is the position vector of the $n$-th element. \eqref{eq:f1} holds if there is no MC among the array elements, which is usually the case when the spacing between adjacent elements is sufficiently large (e.g., more than a couple of wavelengths). Accordingly, the~conventional MC-unaware array response vector is defined as
\begin{equation}\label{eq:a2}
\textbf{a}_0=\left[e^{j\kappa\bold{\hat{d}}\cdot\bold{d}_1},\ldots, e^{j\kappa\bold{\hat{d}}\cdot\bold{d}_n},\ldots, e^{j\kappa\bold{\hat{d}}\cdot\bold{d}_N}\right]^\text{T}.
\end{equation}

Then, the correlation matrix $\textbf{R}_0$ excluding the MC effects is given by
\begin{equation}\label{eq:R2}
\textbf{R}_0=\int_{0}^{\pi}\int_{0}^{\pi}s(\phi,\theta)\textbf{a}_0(\phi,\theta)\textbf{a}_0^\text{H}(\phi,\theta)d\phi d\theta.
\end{equation}

\textcolor{black}{The concrete expression and properties of $\textbf{R}_0$ will be provided in Section~\ref{sec:EV_noMC} to investigate the characteristics of its eigenvalues.}

\subsection{Array Response and Spatial Correlation Including~MC}
\label{sec:SC_noMC}
In \textcolor{black}{a holographic RIS}, the~elements are densely arranged so that MC is usually non-negligible. The~element pattern considering MC can be formulated as~\cite{Sadat04APSS}
\begin{equation}\label{eq:h2}
\tilde{\textcolor{black}{p}}(\phi,\theta)=\textcolor{black}{p}(\phi,\theta)\sum_{m=1}^{N}c_{mn}e^{j\kappa\bold{\hat{d}}\cdot\bold{d}_m}
\end{equation}

\noindent where $c_{mn}$ represents the coupling coefficient between the $m$-th and the $n$-th elements. Namely, the~pattern of the $n$-th element can be modeled as the original element pattern $\textcolor{black}{p}(\phi,\theta)$ times a weighted sum of impact from all the other elements. Thus, the~radiation pattern of the array incorporating MC follows
\begin{equation}\label{eq:f2}
\tilde{f}(\phi,\theta)=\sum_{n=1}^{N}\sum_{m=1}^{N}w_n\textcolor{black}{p}(\phi,\theta)c_{mn}e^{j\kappa\bold{\hat{d}}\cdot\bold{d}_m}.
\end{equation}

The coupling matrix collecting the MC coefficients is written as
\begin{equation}\label{eq:C1}
\textbf{C} = 
\begin{bmatrix}
c_{11} & c_{12} & \cdots & c_{1N} \\
c_{21} & c_{22} & \cdots & c_{2N} \\
\vdots  & \vdots  & \ddots & \vdots  \\
c_{N1} & c_{N2} & \cdots & c_{NN}
\end{bmatrix}.
\end{equation}

When there is no MC among the elements, $\textbf{C}$ reverts to an identity matrix. \textcolor{black}{It can be derived from~\eqref{eq:a2},~\eqref{eq:h2}, and~\eqref{eq:C1} that} the effective array response vector $\textbf{a}$ involving MC is given by
\begin{equation}\label{eq:a1}
\textbf{a}=\textbf{C}^\text{T}\textbf{a}_0
\end{equation}

\noindent in which $\textbf{a}_0$ stands for the original MC-unaware array response vector \textcolor{black}{in~\eqref{eq:a2}}. 

\textcolor{black}{Now, let us look at a crucial metric relevant to the array response---array gain, which is defined herein as the increase in radiation power of an array compared with that of a single element under the same total excitation power. It is well known in the antenna literature~\cite{Uzkov46,Altshuler05TAP,Ivrlac10TCS} that the array gain of an array with closely-spaced elements can grow with the square of the number of elements, and rigorous proof for a linear array was provided in~\cite{Altshuler05TAP} with optimal beamforming in the end-fire direction. For an arbitrary pointing direction $(\phi,\theta)$, the array gain $G_\text{array}$ incorporating MC can be formulated based on \eqref{eq:f2} as}
\begin{equation}\label{eq:AG1}\color{black}
\begin{split}
G_\text{array}(\phi,\theta)&=\frac{\Big|\sum_{n=1}^{N}\sum_{m=1}^{N}w_n\textcolor{black}{p}(\phi,\theta)c_{mn}e^{j\kappa\bold{\hat{d}}\cdot\bold{d}_m}\Big|^2}{|w_0\textcolor{black}{p}(\phi,\theta)|^2}\\
&=\frac{\Big|\sum_{n=1}^{N}\sum_{m=1}^{N}w_nc_{mn}e^{j\kappa\bold{\hat{d}}\cdot\bold{d}_m}\Big|^2}{|w_0|^2}\\
&=\frac{\big|\textbf{a}^\text{T}\textbf{w}\big|^2}{|w_0|^2},~\text{s.t.}~||\textbf{w}||_2=|w_0|
\end{split}
\end{equation}

\noindent \textcolor{black}{where $w_0$ stands for the complex excitation coefficient for a single element such that $|w_0|^2$ represents the total excitation power, $\textbf{a}$ is given in~\eqref{eq:a1}, and~$\textbf{w}$ denotes the beamforming column vector consisting of the complex excitation coefficients $w_n$ for an array. Note that $\textbf{w}$ is inherently a function of the pointing direction $(\phi,\theta)$ due to $\textbf{a}$. Consequently, the~optimal beamforming vector maximizing $G_\text{array}(\phi,\theta)$ is}
\begin{equation}\label{eq:w1}\color{black}
\begin{split}
\textbf{w}_\text{opt}=\zeta\textbf{a}^*=\zeta\textbf{C}^\text{H}\textbf{a}_0^*
\end{split}
\end{equation}

\noindent \textcolor{black}{in which $\zeta$ is a normalization factor equal to $\frac{|w_0|}{||\textbf{C}^\text{H}\textbf{a}_0^*||_2}$ to satisfy the power constraint. Plugging~\eqref{eq:w1} into~\eqref{eq:AG1} produces the maximum array gain at the direction $(\phi,\theta)$}
\begin{equation}\label{eq:AG2}\color{black}
\begin{split}
G_\text{array}(\phi,\theta)_\text{max}=\Big|\textbf{a}_0^\text{T}\textbf{C}\textbf{C}^\text{H}\textbf{a}_0^*\Big|.
\end{split}
\end{equation}

\textcolor{black}{It is straightforward to observe from~\eqref{eq:AG2} that, when excluding MC, i.e.,~when $\textbf{C}$ is an identity matrix, the~maximum array gain is $\big|\textbf{a}_0^\text{T}\textbf{a}_0^*\big|=N$ for an arbitrary direction. When taking MC into account, the~maximum array gain becomes $\big|\textbf{a}_0^\text{T}\textbf{C}\textbf{C}^\text{H}\textbf{a}_0^*\big|$ which is usually larger than $N$ as will be shown later by simulations, and~varies with pointing angles.} The array gain in~\eqref{eq:AG2} of a \textcolor{black}{holographic RIS} is highly promising, since it indicates that, compared with a traditional array with $N$ elements, the~received signal-to-noise ratio (SNR) can be enhanced by up to \textcolor{black}{$\frac{\big|\textbf{a}_0^\text{T}\textbf{C}\textbf{C}^\text{H}\textbf{a}_0^*\big|}{N}$} fold for a fixed transmit power using \textcolor{black}{a holographic RIS} with the same number of elements; or, equivalently, the~transmit power can be scaled down by up to a factor of \textcolor{black}{$\frac{N}{\big|\textbf{a}_0^\text{T}\textbf{C}\textbf{C}^\text{H}\textbf{a}_0^*\big|}$} without compromising the received SNR. \textcolor{black}{The disadvantages of the proposed method in~\eqref{eq:w1} are that the coupling matrix $\textbf{C}$ needs to be known (e.g., through rigorous theoretical analysis or measurements) before conducting the beamforming, and~that the achievable array gain in~\eqref{eq:AG2} might be smaller than $N$ in some corner cases (as will be shown later via simulations) depending on the properties of the coupling matrix $\textbf{C}$ and target pointing angles.}

\textcolor{black}{Based upon the array response vector incorporating MC in~\eqref{eq:a1},} the effective spatial correlation matrix $\textbf{R}$ in~\eqref{eq:R1} is expanded as
\begin{equation}\label{eq:R3}
\begin{split}
\textbf{R}&=\mathbb{E}\left\{\textbf{h}\textbf{h}^\text{H}\right\}\\
&=\int_{0}^{\pi}\int_{0}^{\pi}s(\phi,\theta)\textbf{C}^\text{T}\textbf{a}_0\textbf{a}_0^\text{H}(\phi,\theta)\textbf{C}^*d\phi d\theta\\
&=\textbf{C}^\text{T}\left(\int_{0}^{\pi}\int_{0}^{\pi}s(\phi,\theta)\textbf{a}_0\textbf{a}_0^\text{H}(\phi,\theta)d\phi d\theta\right)\textbf{C}^*\\
&=\textbf{C}^\text{T}\textbf{R}_0\textbf{C}^*
\end{split}
\end{equation}

\noindent which implies that each entry in the effective spatial correlation matrix $\textbf{R}$ is determined by the collective effects of all the entries in the original spatial correlation matrix $\textbf{R}_0$ weighted by the relevant entries in the coupling matrix $\textbf{C}$. \textcolor{black}{The theoretical analysis in~\eqref{eq:w1}--\eqref{eq:R3} will be applied in Section~\ref{sec:EV_MC} to study the performance of the proposed beamforming approach and the influence of MC on the effective spatial correlation of holographic RISs.}


\section{Eigenvalue Distributions Without~MC}
\label{sec:EV_noMC}
Denote the number of elements in \textcolor{black}{a holographic RIS} along the $x$ and $z$ directions as $N_x$ and $N_z$, respectively, i.e.,~the total number of elements $N=N_xN_z$. For~the isotropic scattering environment, the~spatial scattering function in~\eqref{eq:R1} is~\cite{Sun21RISModel}
\begin{equation}\label{eq:s2}
s(\phi,\theta)=\frac{\text{sin}\theta}{2\pi},~\phi\in\big[0,\pi\big],\theta\in\big[0,\pi\big].
\end{equation}

\begin{figure}
	\centering
	\includegraphics[width=10 cm]{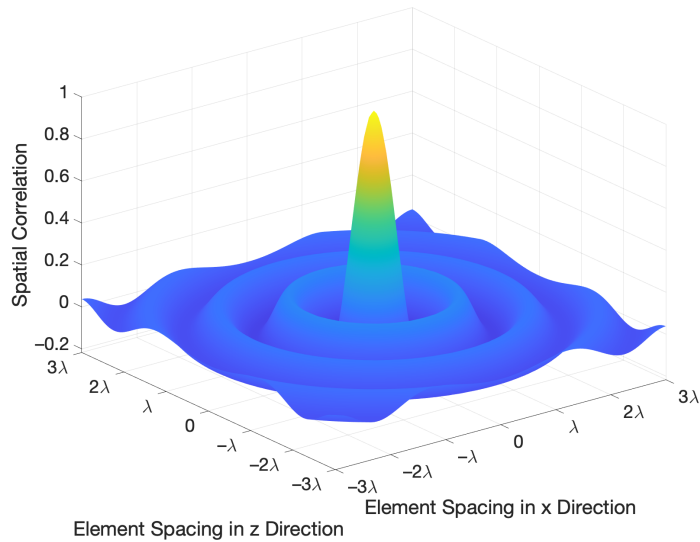}
	\caption{Spatial correlation among the \textcolor{black}{holographic RIS} elements under isotropic scattering, where $\lambda$ denotes the carrier wavelength. \textcolor{black}{Note that the element spacing in $x$ and $z$ directions are expressed in terms of the wavelength $\lambda$, which is directly labeled on the abscissa and ordinate, while the spatial correlation is dimensionless.}}
	\label{fig:sincFunction}
\end{figure}

Substituting~\eqref{eq:s2} into~\eqref{eq:R2} \textcolor{black}{in Section~\ref{sec:SC_noMC_theory}} yields the spatial correlation matrix $\textbf{R}_0$ \textcolor{black}{of the holographic RIS} under isotropic scattering. As~proven in~\cite{Bjornson20WCL,Sun21RISModel}, $\textbf{R}_0$ can be characterized by a sinc function as follows:
\begin{equation}\label{eq:R5}
\begin{split}
[\textbf{R}_0]_{n_1,n_2}=\text{sinc}\bigg(\frac{2||\textbf{d}_{n_1}-\textbf{d}_{n_2}||_2}{\lambda}\bigg),~n_1,n_2=1,\ldots,N
\end{split}
\end{equation}

\noindent where $\text{sinc}(x)\triangleq\frac{\text{sin}(\pi x)}{\pi x}$ is the sinc function, $\textbf{d}_{n_1}$ and $\textbf{d}_{n_2}$ denote the coordinates of the $n_1$-th and $n_2$-th elements \textcolor{black}{in the holographic RIS}, respectively. The~behavior of small-scale spatial correlation is depicted in Fig.~\ref{fig:sincFunction} for element spacing up to three times the wavelength $\lambda$. It is observed from (\ref{eq:R5}) and Fig.~\ref{fig:sincFunction} that the spatial correlation is minimal only for some element spacing, instead of between any two elements, thus the classical independent and identically distributed (i.i.d.) 
Rayleigh fading model is not applicable in such a system~\cite{Pizzo20JSAC,Bjornson20WCL,Sun21RISModel}.

\begin{figure}
	\centering
	\includegraphics[width=0.7\columnwidth]{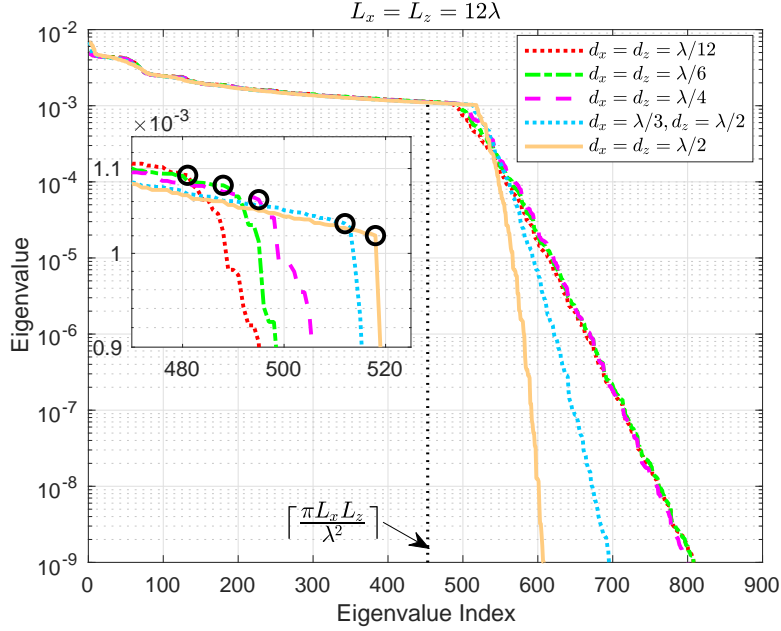}
	\caption{Eigenvalue versus eigenvalue index of $\textbf{R}_0/N$ in non-increasing order for various element spacing $d_x$ and $d_z$ with $L_x=L_z=12\lambda$. In addition, the asymptotic spatial degrees of freedom (DoF) $\lceil\frac{\pi L_xL_z}{\lambda^2}\rceil$ derived in~{\cite{Pizzo22TWC}} are depicted for $\text{min}(L_x,L_z)/\lambda\to\infty$. \textcolor{black}{Note that both the eigenvalue and its index are dimensionless.} Each dark circle in the inset represents the inflection point where the eigenvalues start to drop rapidly.}
	\label{fig:EV_12}	
\end{figure}

Fig.~\ref{fig:EV_12} portrays the eigenvalues of $\textbf{R}_0/N$ in non-increasing order for various $N$, or~equivalently various element spacing, with~$L_x=L_z=12\lambda$. In addition, the~dotted vertical line represents the asymptotic spatial DoF $\lceil\frac{\pi L_xL_z}{\lambda^2}\rceil$ for $\text{min}(L_x,L_z)/\lambda\to\infty$ \cite{Pizzo22TWC,Franceschetti15TIT}. A~few key remarks can be drawn from Fig.~\ref{fig:EV_12}: First, while the popular i.i.d. Rayleigh fading channel has almost identical eigenvalues whose amount equals the number of antenna elements deployed, the~correlated channel herein has uneven and fewer dominant eigenvalues (those larger than the ones highlighted by the dark circles in the inset of Fig.~\ref{fig:EV_12}) and smaller rank. Second, the~accuracy of the spatial DoF $\lceil\frac{\pi L_xL_z}{\lambda^2}\rceil$ increases with the element density. More importantly, the~number of dominant eigenvalues, highlighted by the dark circles in the inset of Fig.~\ref{fig:EV_12}, declines as the total number of elements $N$ increases, which seems counter-intuitive. Therefore, in~the following subsections, we will dig into the underlying causes of the aforementioned uncommon phenomenon \textcolor{black}{of seemingly reduced spatial DoF with an increased number of elements in a holographic RIS}. 

\vspace{-6pt}

\unskip

\subsection{Relationship between Eigenvalues and Power~Spectrum}
Note that the spatial correlation matrix $\textbf{R}_0$ of the 2D \textcolor{black}{holographic RIS} in~\eqref{eq:R5} can be formulated as
\begin{equation}\label{eq:R6}
\textbf{R}_0= 
\begin{bmatrix}
	\textbf{B}_{0} & \textbf{B}_{1} & \textbf{B}_{2}&\cdots & \textbf{B}_{N_z-1}\\
\textbf{B}_{-1} & \textbf{B}_{0} &  \textbf{B}_{1}&\cdots & \textbf{B}_{N_z-2}\\
\textbf{B}_{-2} & \textbf{B}_{-1} &  \textbf{B}_{0}&\cdots & \textbf{B}_{N_z-3}\\
\vdots & \vdots & \vdots & \ddots & \vdots\\
\textbf{B}_{1-N_z} & \textbf{B}_{2-N_z} &\textbf{B}_{3-N_z}& \cdots & \textbf{B}_{0}\\
\end{bmatrix}
\end{equation}

\noindent where each block $\textbf{B}_{m}\in\mathbb{C}^{N_x\times N_x}$, $|m|<N_z$, is a symmetric Toeplitz matrix by itself, and~$\textbf{B}_{-m}=\textbf{B}_{m}^T$, i.e.,~the entire matrix $\textbf{R}_0$ is symmetric as well. Moreover, the~matrix blocks $\textbf{B}_{m}$'s along each diagonal of $\textbf{R}_0$ are identical. Therefore, the~spatial correlation matrix $\textbf{R}_0$ has a symmetric BTTB structure under isotropic scattering, thus we resort to the relevant theory of the asymptotic distribution of eigenvalues of BTTB forms to investigate the properties of the eigenvalues of $\textbf{R}_0/N$. 

Denote the $(u,v)$th entry of $\textbf{B}_{m}$ by $b_{l,m}=b_{u-v,m}$, $u,v=1,\ldots,N_x$, and define the function
\begin{equation}
g(\Delta_x,\Delta_z) = \text{sinc}\left(\frac{2\sqrt{\Delta_x^2+\Delta_z^2}}{\lambda}\right)
\label{eq:g1}
\end{equation}

\noindent where $\Delta_x$ and $\Delta_z$ are the spacing along the $x$- and $z$-axes, respectively, between~a pair of spatial points of interest in the $xoz$ plane. $\{b_{l,m}\}$ can thus be regarded as a finite truncated bi-sequence generated from $g(\Delta_x,\Delta_z)$, and~more precisely,
\begin{equation}
b_{l,m}=\text{sinc}\left(\frac{2\sqrt{\left(\frac{lL_x}{N_x}\right)^2+\left(\frac{mL_z}{N_z}\right)^2}}{\lambda}\right).
\label{eq:b1}
\end{equation}

Let $G(\omega_x,\omega_z)$ be the 2D Fourier transform of $\{b_{l,m}\}$ given by
\begin{equation}
\begin{aligned}
&G(\omega_x,\omega_z)=\frac{1}{N}\sum_{l=-(N_x-1)}^{N_x-1}\sum_{m=-(N_z-1)}^{N_z-1}b_{l,m}e^{-j(l\omega_x+m\omega_z)},\\
&\omega_x=-\frac{(N_x-1)\pi}{N_x},-\frac{(N_x-3)\pi}{N_x},\ldots,\frac{(N_x-1)\pi}{N_x},\\
&\omega_z=-\frac{(N_z-1)\pi}{N_z},-\frac{(N_z-3)\pi}{N_z},\ldots,\frac{(N_z-1)\pi}{N_z}.
\end{aligned}
\label{eq:G1}
\end{equation}

Since the double-index sequence $\{b_{l,m}\}$ consists of spatial samples of the continuous sinc function $g(\Delta_x,\Delta_z)$ in~\eqref{eq:g1}, $G(\omega_x,\omega_z)$ in~\eqref{eq:G1} can be looked upon as the power spectrum of the discretized and truncated spatial correlation function. According to the properties of BTTB matrices, the~eigenvalues of $\textbf{R}_0/N$ behave asymptotically the same as the \textcolor{black}{spectral} sampling points of $G(\omega_x,\omega_z)$ as $N\to\infty$, if~the double-index sequence formed by $b_{l,m}$ is absolutely summable~\cite{Chan93SJNA,Bose98TIT}, i.e.,
\begin{equation}
\lim_{N_x,N_z\to\infty}\sum_{l=-(N_x-1)}^{N_x-1}\sum_{m=-(N_z-1)}^{N_z-1}\left|b_{l,m}\right|\leq\text{Constant}<\infty.
\label{b2}
\end{equation}

The condition in~\eqref{b2} is indeed satisfied as shown by~\eqref{b3}, 
\begin{equation}
	\begin{aligned}
	&\lim_{N_x,N_z\to\infty}\sum_{l=-(N_x-1)}^{N_x-1}\sum_{m=-(N_z-1)}^{N_z-1}\left|b_{l,m}\right|\\
	=&\lim_{N_x,N_z\to\infty}\sum_{l=-(N_x-1)}^{N_x-1}\sum_{m=-(N_z-1)}^{N_z-1}\left|\text{sinc}\left(\frac{2\sqrt{\left(\frac{lL_x}{N_x}\right)^2+\left(\frac{mL_z}{N_z}\right)^2}}{\lambda}\right)\right|\\
	=&\int_{-1}^{1}\int_{-1}^{1}\left|\text{sinc}\left(\frac{2\sqrt{\left(\varpi L_x\right)^2+\left(\varsigma L_z\right)^2}}{\lambda}\right)\right|d\varpi d\varsigma\\
	\leq&\int_{-1}^{1}\int_{-1}^{1}1d\varpi d\varsigma=4<\infty\\
	\end{aligned}
	\label{b3}
	\end{equation}
hence \textcolor{black}{the eigenvalues of $\textbf{R}_0/N$ and the spectral sampling points of $G(\omega_x,\omega_z)$ in~\eqref{eq:G1} are asymptotically equally distributed. Consequently, insights on the eigenvalues of $\textbf{R}_0/N$ can be drawn via the investigation of $G(\omega_x,\omega_z)$.} 
In what follows, we explore how the power spectrum $G(\omega_x,\omega_z)$ changes with the element spacing of the \textcolor{black}{holographic RIS}, or~equivalently, the~spatial sampling frequency, in~order to explain the unconventional observation of seemingly lower spatial DoF with growing numbers of elements in \textcolor{black}{a holographic RIS} as manifest in Fig.~\ref{fig:EV_12}.

\subsection{Analysis on Eigenvalues via Power~Spectrum}
When the element spacing in \textcolor{black}{a holographic RIS} is $\eta_x\lambda$ and $\eta_z\lambda$ ($\eta_x,\eta_z>0$) along the $x$ and $z$ directions, respectively, and~$L_x=\beta_x\lambda$, $L_z=\beta_z\lambda$ ($\beta_x,\beta_z>0$), we obtain $N_x=\beta_x/\eta_x+1$, $N_z=\beta_z/\eta_z+1$. The~power spectrum $G(\omega_x,\omega_z)$ in~\eqref{eq:G1} can be recast as~\eqref{eq:G2}, 
\vspace{-8pt}
\begin{equation}
	\begin{aligned}
	G(\kappa_x,\kappa_z)&=\frac{1}{N}\sum_{l=-\frac{\beta_x}{\eta_x}}^{\frac{\beta_x}{\eta_x}}\sum_{m=-\frac{\beta_z}{\eta_z}}^{\frac{\beta_z}{\eta_z}}\text{sinc}\left(2\sqrt{\left(\frac{l\eta_x(N_x-1)}{N_x}\right)^2+\left(\frac{m\eta_z(N_z-1)}{N_z}\right)^2}\right)e^{-j\lambda(l\eta_x\kappa_x+m\eta_z\kappa_z)},\\
	\kappa_x&=-\frac{(N_x-1)\kappa}{2N_x\eta_x},-\frac{(N_x-3)\kappa}{2N_x\eta_x},\ldots,\frac{(N_x-1)\kappa}{2N_x\eta_x},\\
	\kappa_z&=-\frac{(N_z-1)\kappa}{2N_z\eta_z},-\frac{(N_z-3)\kappa}{2N_z\eta_z},\ldots,\frac{(N_z-1)\kappa}{2N_z\eta_z}
	\end{aligned}
	\label{eq:G2}
	\end{equation}
in which $\kappa=2\pi/\lambda$ denotes the wavenumber; $\kappa_x$ and $\kappa_z$ represent the wavenumber along the $x$ and $z$ directions, respectively. The wavenumber along the positive $y$ direction in Fig.~\ref{fig:SystemModel} is defined as
\begin{equation}
\varkappa(\kappa_x,\kappa_z)=\sqrt{\kappa^2-\left(\kappa_x^2+\kappa_z^2\right)}
\end{equation}

\noindent and the wave vector is
\begin{equation}
 \boldsymbol{\kappa}=\kappa_x\bold{\hat{x}}+\varkappa(\kappa_x,\kappa_z)\bold{\hat{y}}+\kappa_z\bold{\hat{z}}
 \end{equation}
 
 \noindent where $\bf{\hat{x}}$, $\bf{\hat{y}}$, and~$\bf{\hat{z}}$ denote the unit vector along the $x$-, $y$-, and~$z$-axis, respectively. It is noteworthy that a real-valued $\varkappa(\kappa_x,\kappa_z)$ corresponds to a wave propagating along the $y$-direction, while an imaginary-valued $\varkappa(\kappa_x,\kappa_z)$ indicates an evanescent wave that usually exists around the surface of an object and decays exponentially in space~\cite{Pizzo20JSAC,Sun_JOSA}. 

\begin{figure}
	\centering
	\includegraphics[width=0.7\columnwidth]{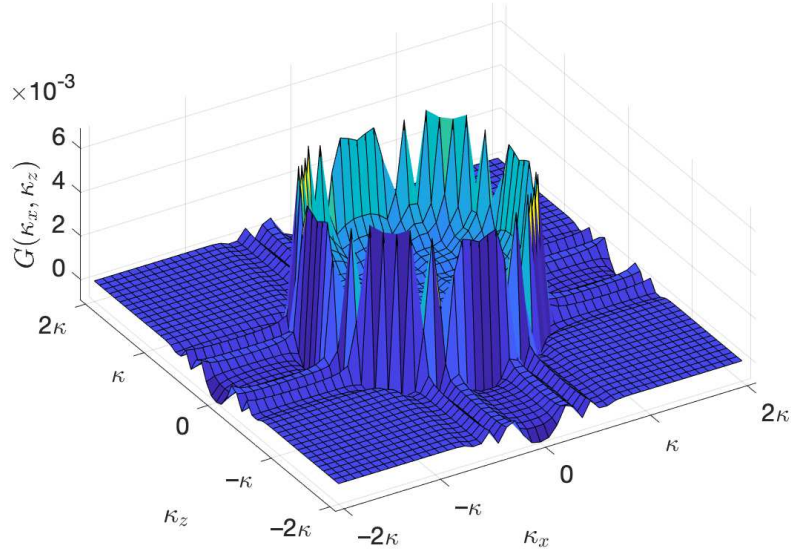}
	\caption{Fourier transform $G(\kappa_x,\kappa_z)$ in~\eqref{eq:G2} of the truncated and discretized spatial correlation function $g(\Delta_x,\Delta_z)$ in~\eqref{eq:g1} with $L_x=L_z=12\lambda$ and $d_x=d_z=\lambda/3$. \textcolor{black}{Note that $\kappa_x$ and $\kappa_z$ are expressed in terms of the wavenumber $\kappa$, which is directly labeled on the abscissa and ordinate, while $G(\kappa_x,\kappa_z)$ is dimensionless since $g(\Delta_x,\Delta_z)$ is dimensionless.}}
	\label{fig:ft_3D}	
\end{figure}

The formulation in~\eqref{eq:G2} allows us to study the power spectrum $G(\kappa_x,\kappa_z)$ as a function of the wavenumbers $\kappa_x$ and $\kappa_z$, so as to gain more insight on the influence of the spatial sampling frequency on $G(\kappa_x,\kappa_z)$, and,~equivalently, on~the eigenvalues of the normalized spatial correlation matrix $\textbf{R}_0/N$, thanks to the time-frequency and space-wavenumber duality~\cite{Pizzo22TWC}. For~a continuous and infinitely large \textcolor{black}{holographic RIS} aperture, $G(\kappa_x,\kappa_z)$ in~\eqref{eq:G2} approaches the following distribution~\cite{Bracewell99Book}:
\begin{equation}
\begin{aligned}
G^{(\infty)}(\kappa_x,\kappa_z)=\frac{\Pi\left(\frac{\sqrt{\kappa_x^2+\kappa_z^2}}{2\kappa}\right)}{\frac{\kappa}{2\pi}\sqrt{\kappa^2-(\kappa_x^2+\kappa_z^2)}}
\end{aligned}
\label{eq:G3}
\end{equation}

\noindent where $\Pi(\cdot)$ is the rectangle function. As~implied by~\eqref{eq:G3}, $G^{(\infty)}(\kappa_x,\kappa_z)$ assumes a bowl-like shape for a given $\kappa$, which increases monotonically with $\sqrt{\kappa_x^2+\kappa_z^2}$ inside the region
\begin{equation}
\mathcal{D}(\kappa)=\left\{(\kappa_x,\kappa_z)\in\mathbb{R}^2:\kappa_x^2+\kappa_z^2\leq\kappa^2\right\}
\label{eq:D}
\end{equation}

\noindent achieving the maximum value when $\sqrt{\kappa_x^2+\kappa_z^2}=\kappa$, and~then transits abruptly to zero outside $\mathcal{D}(\kappa)$. In~practical implementations, however, \textcolor{black}{a holographic RIS} often has a finite aperture size and is composed of discrete elements, which is equivalent to \textcolor{black}{spatially} \textit{truncating} and \textit{sampling} an originally infinite and continuous \textcolor{black}{holographic RIS} aperture. Truncating can be thought of as \textcolor{black}{applying} a windowing function to \textcolor{black}{an originally infinitely-long} signal \textcolor{black}{in the spatial domain}, which will cause the signal to appear outside $\mathcal{D}(\kappa)$ in the wavenumber domain \textcolor{black}{after performing the Fourier transform}, as~displayed in Fig.~\ref{fig:ft_3D}. Regarding spatial sampling, it is known from time-frequency-domain signal processing that a finer sampling granularity in the time domain yields a higher resolution in the (traditional) frequency domain. Analogously, for~\textcolor{black}{a holographic RIS}, a~smaller \textcolor{black}{spatial} sampling interval, which entails a smaller element spacing, gives rise to a higher resolution in the spatial frequency (i.e., wavenumber) domain. Accordingly, taking the $x$ direction as an example, the~highest absolute value of the resolvable wavenumber $\kappa_x$, namely $\frac{(N_x-1)\kappa}{2N_x\eta_x}$, is inversely proportional to the element spacing $\eta_x$ in~\eqref{eq:G2}. Specifically, if~$N_x\gg1$, for~the most common element spacing of half a wavelength (i.e., $\eta_x=1/2$), the~highest resolvable wavenumber is $\kappa$ as expected, and~it increases to $\frac{\kappa}{2\eta_x}$ as $\eta_x$ decreases. 
\unskip

\begin{figure}
	\centering
	\subfloat[\textcolor{black}{$d_x=d_z=\lambda/2$}]{\includegraphics[width=0.43\textwidth]{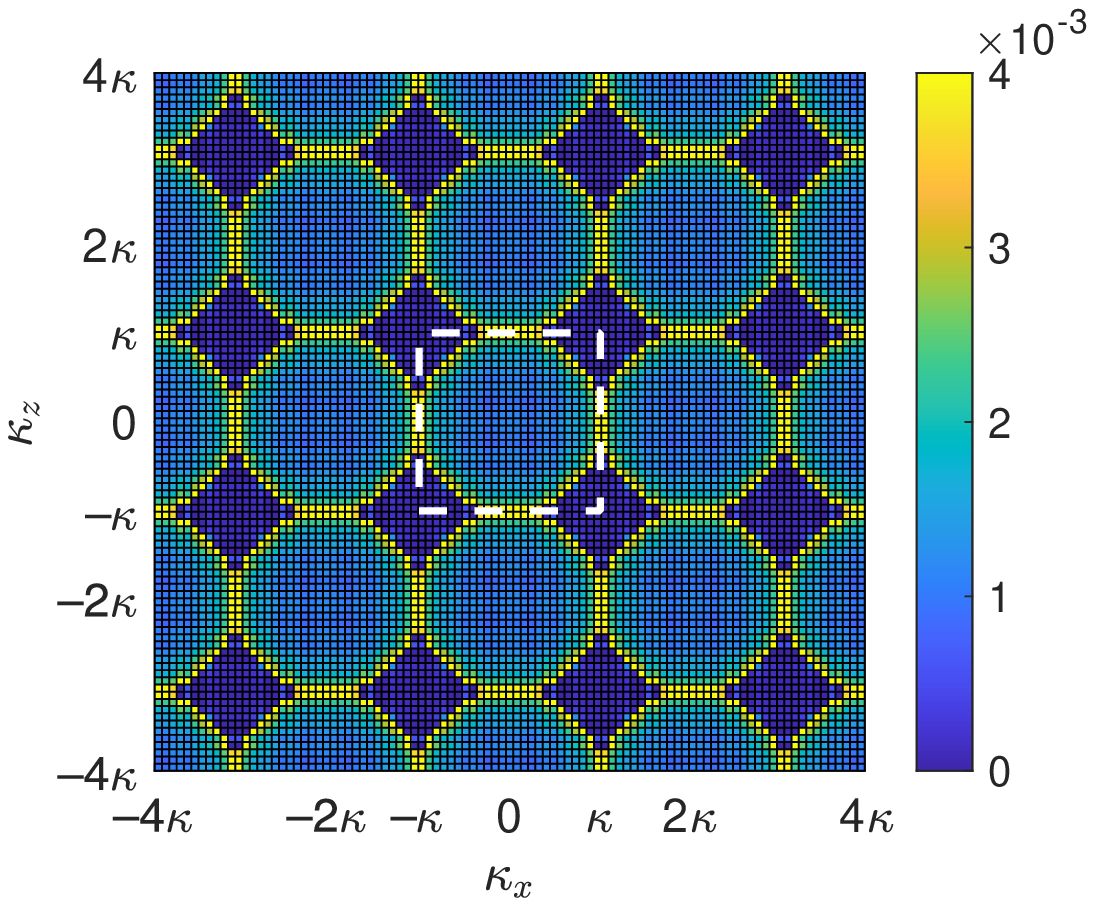}}
	\subfloat[\textcolor{black}{$d_x=d_z=\lambda/2$}]{\includegraphics[width=0.43\textwidth]{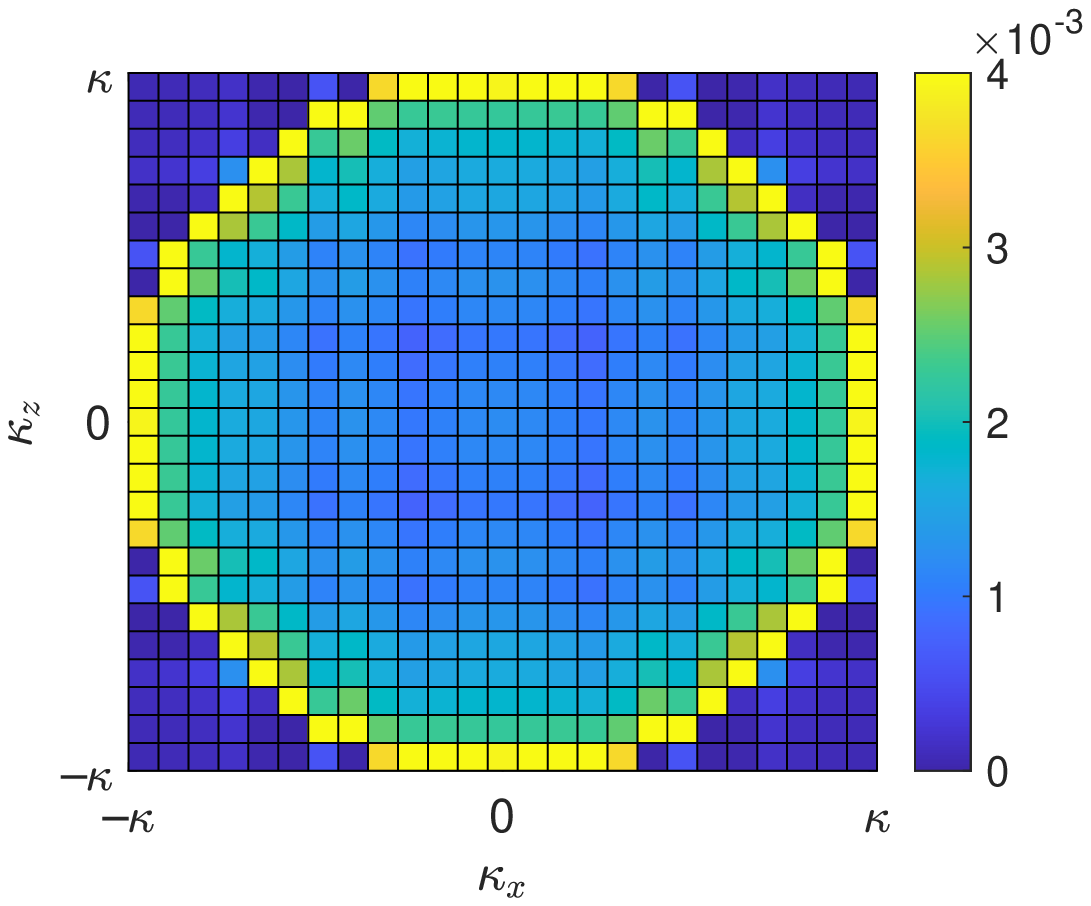}}\vspace{2.6pt}\\
	\subfloat[\textcolor{black}{$d_x=d_z=\lambda/4$}]{\includegraphics[width=0.43\textwidth]{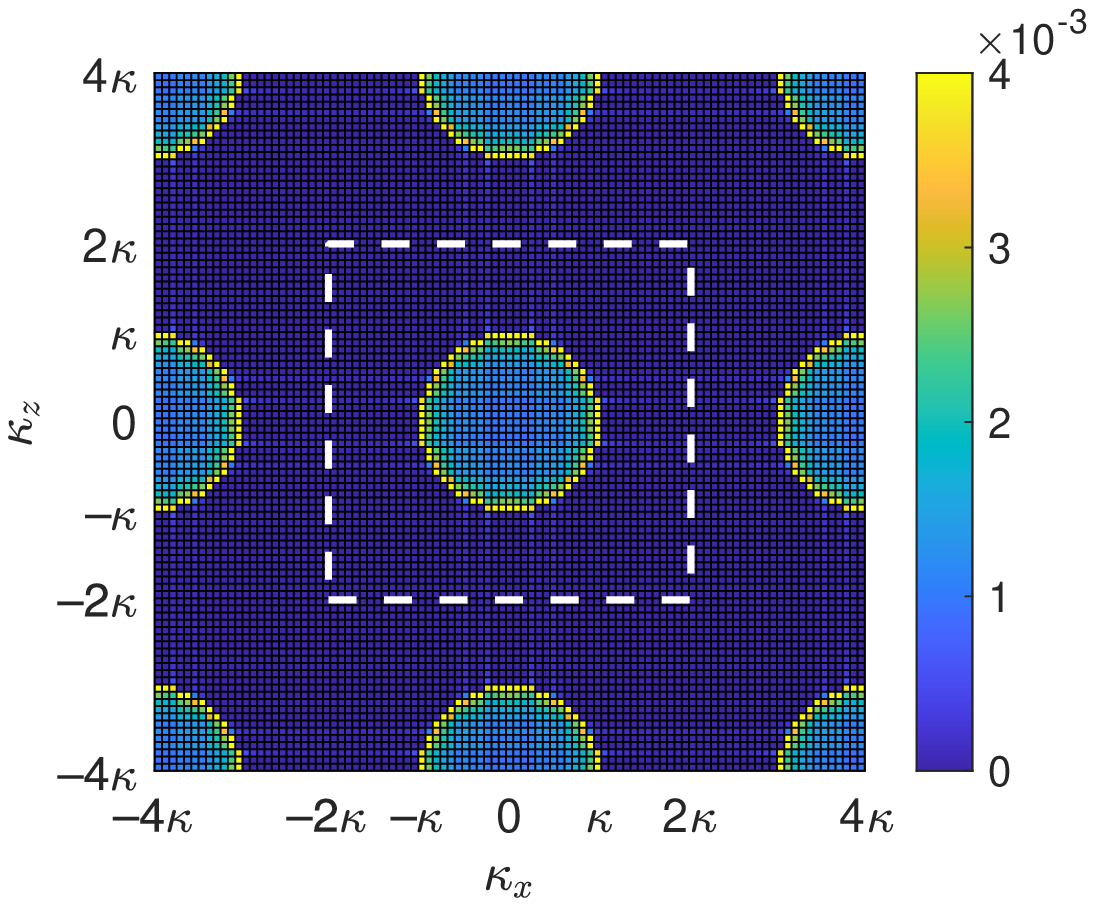}}
	\subfloat[\textcolor{black}{$d_x=d_z=\lambda/4$}]{\includegraphics[width=0.43\textwidth]{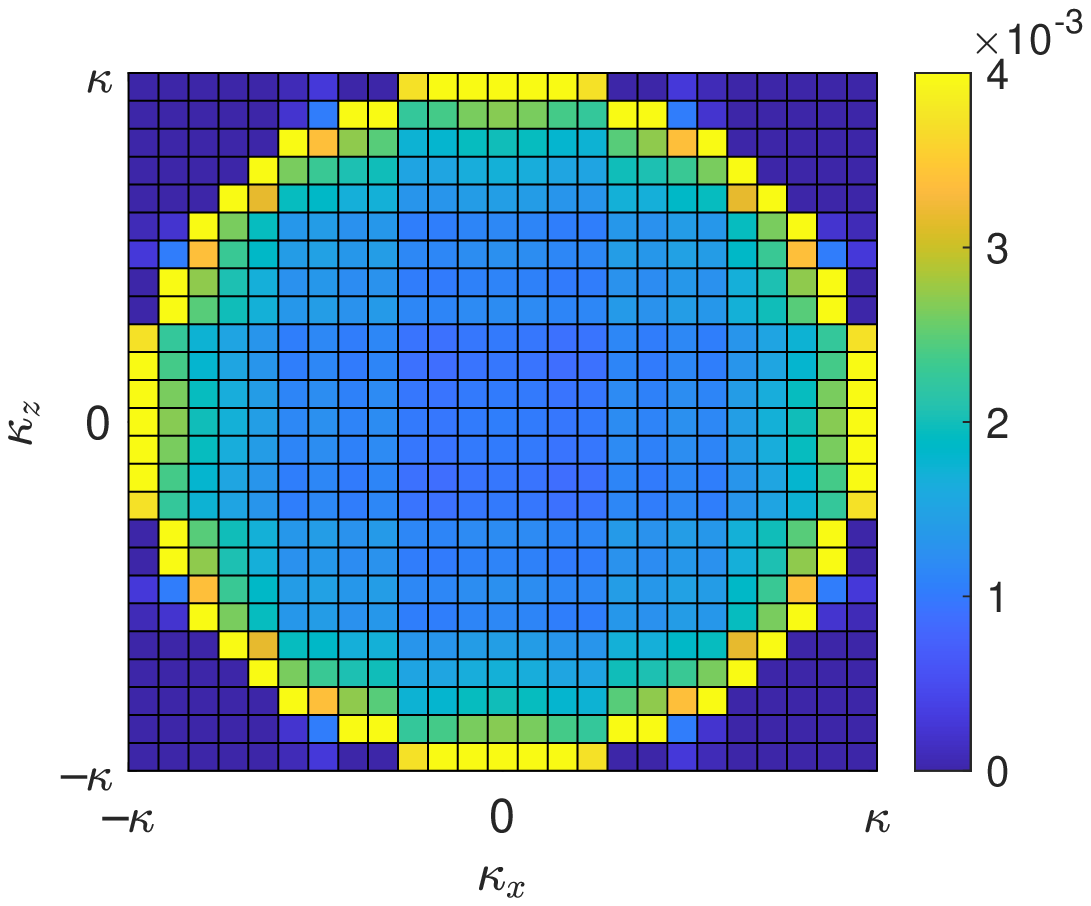}}\vspace{2.6pt}\\
	\subfloat[\textcolor{black}{$d_x=d_z=\lambda/8$}]{\includegraphics[width=0.43\textwidth]{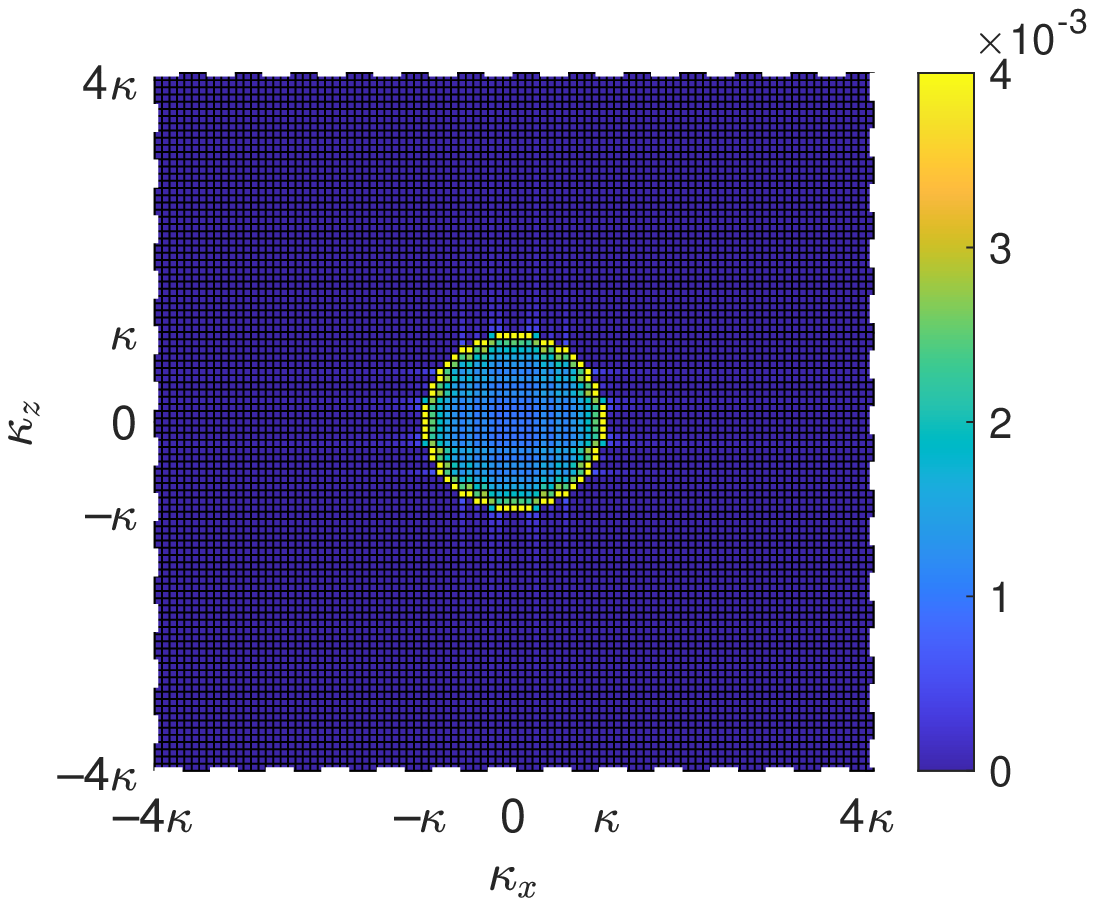}}
	\subfloat[\textcolor{black}{$d_x=d_z=\lambda/8$}]{\includegraphics[width=0.43\textwidth]{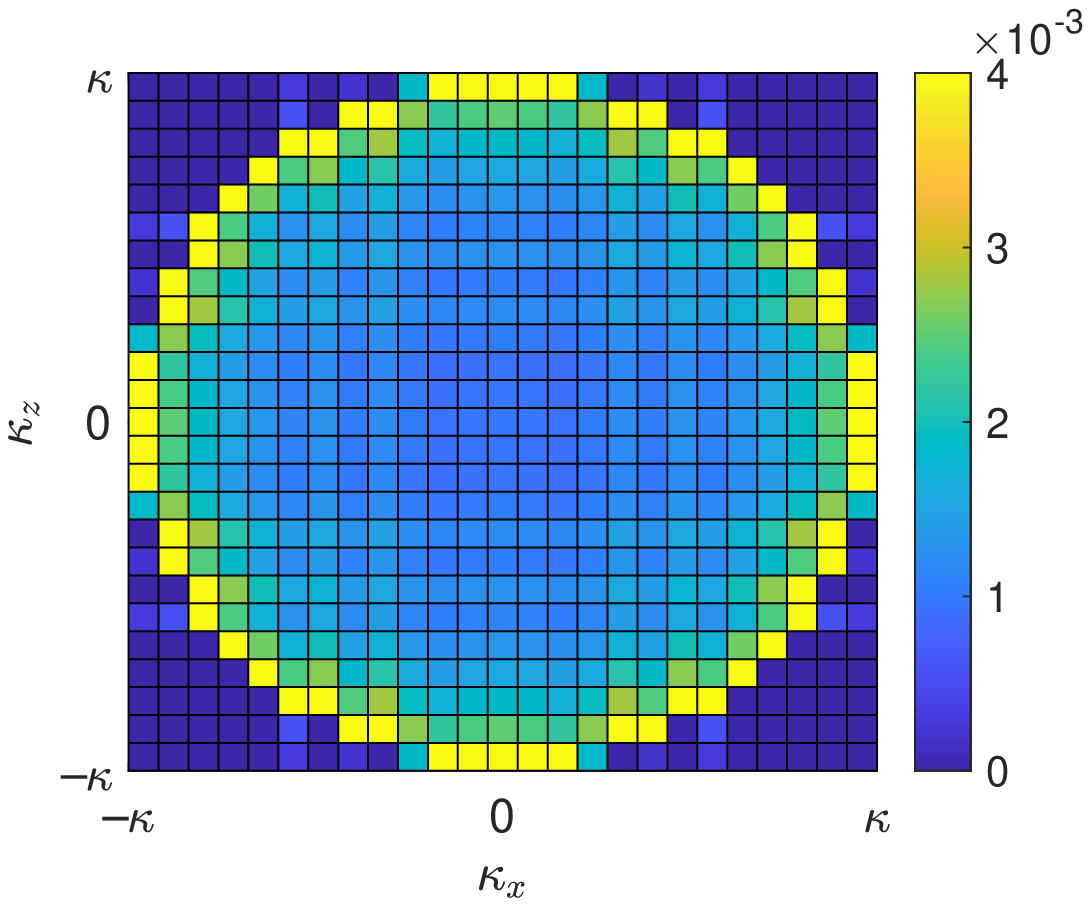}}\vspace{2pt}
	\caption{Overview (\textbf{a},\textbf{c},\textbf{e}) and zoom-in views (\textbf{b},\textbf{d},\textbf{f}) of the Fourier transform $G(\kappa_x,\kappa_z)$ in~\eqref{eq:G2} of the truncated and discretized spatial correlation function $g(\Delta_x,\Delta_z)$ in~\eqref{eq:g1}. The~colorbars represent the values of $G(\kappa_x,\kappa_z)$, and~the white dashed frames in (\textbf{a},\textbf{c},\textbf{e}) outline the regions within which the wavenumbers are resolvable for the corresponding \textcolor{black}{holographic RIS}. \textcolor{black}{Note that $\kappa_x$ and $\kappa_z$ are expressed in terms of the wavenumber $\kappa$ which is directly labeled on the abscissa and ordinate in \mbox{each subfigure.}}}
	\label{fig:ft}
\end{figure}

To examine the impact of the spatial sampling interval on the power spectrum $G(\kappa_x,\kappa_z)$, we perform numerical simulations with a series of $\eta_x$ and $\eta_z$ values, and~the corresponding results are shown in Fig.~\ref{fig:ft}. Figs.~\ref{fig:ft}(a),(c),(e) illustrate the power spectrum $G(\kappa_x,\kappa_z)$ evaluated at $\eta_x=\eta_z=1/2$, $\eta_x=\eta_z=1/4$, and~$\eta_x=\eta_z=1/8$, respectively. For~fair comparison, $\kappa_x$ and $\kappa_z$ range from $-4\kappa$ to $4\kappa$ in all of the three subfigures, while the actual wavenumber regime that the corresponding \textcolor{black}{holographic RIS} can resolve is within the white dashed frame in each of the three subfigures. Figs.~\ref{fig:ft}(b),(d),(f) are the zoom-in views of Figs.~\ref{fig:ft}(a),(c),(e), respectively, where $|\kappa_x|\leq\kappa$ and $|\kappa_z|\leq\kappa$. Since $b_{l,m}$ in~\eqref{eq:b1} can be treated as an aperiodic discrete-space signal, its Fourier transform (i.e., spectrum in the wavenumber domain) $G(\kappa_x,\kappa_z)$ in~\eqref{eq:G2} is periodic with periodicities of $\kappa_x/\eta_x$ and $\kappa_z/\eta_z$ along the $x$- and $z$-directions, respectively, as~demonstrated in Figs.~\ref{fig:ft}(a),(c),(e). Therefore, the~spectrum in the wavenumber domain is actually superpositions of the original spectrum and its replicas, and~the larger the element spacing, the more serious the aliasing is. Due to the presence of spectrum leakage outside the region $\mathcal{D}(\kappa)$ in~\eqref{eq:D} as shown by Fig.~\ref{fig:ft_3D}, the~spectrum aliasing magnifies some of the spectrum values around the periphery of $\mathcal{D}(\kappa)$, and this magnification is more evident for larger element spacing, as~evidenced in Figs.~\ref{fig:ft}(b),(d),(f). Recall that the eigenvalues of $\textbf{R}_0/N$ are asymptotically equally distributed with the corresponding power spectrum $G(\kappa_x,\kappa_z)$, hence the number of significant eigenvalues appears larger for greater element spacing, which explains the observation highlighted by the dark circles in Fig.~\ref{fig:EV_12}. In~other words, the~number of eigenvalues lying in the far-field propagating wave region $\mathcal{D}(\kappa)$ in~\eqref{eq:D} actually does not change with the element spacing, the~specious more dominant eigenvalues and hence higher spatial DoF for a smaller number of elements are contributed by evanescent waves outside the region $\mathcal{D}(\kappa)$. 

Evanescent waves are usually localized in the reactive near field of an array. They contain the high-spatial-frequency components of an object, and~normally do not contribute to the far-field channel capacity, but have a huge potential
in near-field communication, sensing, and~power transfer~\cite{Yin13TBCS,Bjornson19DSP}. Moreover, diverse approaches have been put forward to convert evanescent waves to propagating waves that can be radiated to the far field, e.g.,~via simple metal strip gratings, metasurfaces, or~randomly distributed metal wires~\cite{Lerosey07Science,Memarian12TMTT,Li20LPR}. Overall, \textcolor{black}{holographic RISs}, along with the associated propagating and evanescent waves, can find expansive potential applications in future communications, sensing, and~related~realms. 

\section{Eigenvalue Distributions with~MC}
\label{sec:EV_MC}
In this section, we investigate eigenvalue distributions taking into account MC in \textcolor{black}{holographic RISs}. 

\subsection{Coupling~Matrix}
The impedance matrix $\textbf{Z}$ of an antenna array can be expressed as
\begin{equation}\label{eq:Z1}
\textbf{Z} = 
\begin{bmatrix}
z_\text{A} & z_{12} & \cdots & z_{1N} \\
z_{21} & z_\text{A} & \cdots & z_{2N} \\
\vdots  & \vdots  & \ddots & \vdots  \\
z_{N1} & z_{N2} & \cdots & z_\text{A} 
\end{bmatrix}
\end{equation}

\noindent with $z_\text{A}$ denoting the antenna impedance, and~$z_{mn}$ the mutual impedance between the $m$-th and $n$-th elements. Given $\textbf{Z}$, the~(unnormalized) coupling matrix of the array at the Tx side can be calculated based on circuit theory as~\cite{Janaswamy02AWPL,Chen18Access}
\begin{equation}\label{eq:C2}
\textbf{C}_\text{T}^\prime=\textbf{Z}\big(\textbf{Z}+z_\text{S}\textbf{I}_N\big)^{-1}
\end{equation}

\noindent where $z_\text{S}$ is the source impedance. If~there is no MC between the Tx elements, then $\textbf{Z}$ is diagonal with entries $z_\text{A}$, hence $\textbf{C}_\text{T}^\prime$ is diagonal as well with entries $[\textbf{C}_\text{T}^\prime]_{nn}=z_\text{A}/(z_\text{A}+z_\text{S})$. Consequently, we normalize the coupling matrix by $z_\text{A}/(z_\text{A}+z_\text{S})$ to obtain
\begin{equation}\label{eq:C3}
\textbf{C}_\text{T}=\left(1+z_\text{S}/z_\text{A}\right)\textbf{Z}\big(\textbf{Z}+z_\text{S}\textbf{I}_N\big)^{-1}.
\end{equation}

On the other hand, the~(unnormalized) coupling matrix of the array at the Rx side is given by~\cite{Janaswamy02AWPL,Chen18Access}
\begin{equation}\label{eq:C4}
\textbf{C}_\text{R}^\prime=z_\text{L}\textbf{I}_N\big(\textbf{Z}+z_\text{L}\textbf{I}_N\big)^{-1}
\end{equation}

\noindent where $z_\text{L}$ is the load or termination impedance. Analogous to the Tx side, if~no MC exists between the Rx elements, then $\textbf{C}_\text{R}^\prime$ is diagonal with entries $[\textbf{C}_\text{R}^\prime]_{nn}=z_\text{L}/(z_\text{A}+z_\text{L})$, such that the normalized coupling matrix is
\begin{equation}\label{eq:C5}
\textbf{C}_\text{R}=\left(z_\text{A}+z_\text{L}\right)\big(\textbf{Z}+z_\text{L}\textbf{I}_N\big)^{-1}.
\end{equation}

\textcolor{black}{Given the coupling matrix, the~effective spatial correlation matrix $\textbf{R}$ for a holographic RIS can now be computed by applying the theoretical analysis in Section~\ref{sec:sysModel}. The~steps for deriving $\textbf{R}$ are summarized below:}
\begin{itemize}\color{black}
	\item Step 1: Calculate the conventional (i.e., MC-unaware) array response vector $\textbf{a}_0$ in~\eqref{eq:a2} based on the geometry and element topology of the RIS;
	\item Step 2: Obtain the spatial scattering function $s(\phi,\theta)$ in~\eqref{eq:s1} based on theoretical analysis or measurements;
	\item Step 3: Calculate the spatial correlation matrix $\textbf{R}_0$ that excludes MC according to~\eqref{eq:R2};
	\item Step 4: Obtain the impedance matrix $\textbf{Z}$ in~\eqref{eq:Z1} for the elements in a holographic RIS based on theoretical analysis or measurements;
	\item Step 5: Calculate the coupling matrix at the Tx and Rx, $\textbf{C}_\text{T}$ and $\textbf{C}_\text{R}$, according to~\eqref{eq:C3} and~\eqref{eq:C5}, respectively;
	\item Step 6: Calculate the effective spatial correlation matrix $\textbf{R}$ according to~\eqref{eq:R3} using $\textbf{R}_0$ from Step 3 and $\textbf{C}_\text{T}$ ($\textbf{C}_\text{R}$) from Step 5 for the Tx (Rx).
\end{itemize}	

\subsection{Inter-Element Correlation/Coupling Strength~Indicator}
To facilitate the quantitative description of the physical MC or correlation strength represented by a coupling or correlation matrix $\textbf{Q}\in\mathbb{C}^{N\times N}$, we propound a new metric called \textit{ICSI} defined as
\begin{equation}\label{eq:ICSI1}
\begin{split}
\text{ICSI}&=\frac{1}{N}\sum_{n=1}^{N}\frac{\frac{1}{N-1}\sum_{m=1,m\neq n}^{N}\big|[\textbf{Q}]_{n,m}\big|}{\big|[\textbf{Q}]_{n,n}\big|}\\
&=\frac{1}{N(N-1)}\sum_{n=1}^{N}\sum_{m=1,m\neq n}^{N}\frac{\big|[\textbf{Q}]_{n,m}\big|}{\big|[\textbf{Q}]_{n,n}\big|}
\end{split}
\end{equation}

\noindent where $\frac{\big|[\textbf{Q}]_{n,m}\big|}{\big|[\textbf{Q}]_{n,n}\big|}$ represents the inter-element correlation/coupling strength between the $n$-th and $m$-th ($m\neq n$) elements normalized by the self-correlation/coupling magnitude of the $n$-th element to ease the comparison between different matrices, then the normalized inter-element correlation/coupling magnitude is averaged over all pairs of distinct array elements to arrive at the mean normalized inter-element correlation/coupling strength, i.e.,~the ICSI. The~value of ICSI lies between 0 and 1, which indicate no MC/correlation and maximum MC/correlation between different array elements, and~roughly correspond to the highest and lowest level of orthogonality among the columns/rows of the correlation or coupling matrix $\textbf{Q}$, respectively. A~large value of ICSI entails intense correlation/coupling between distinct array elements on average, which is likely to happen when the correlation/coupling between some pairs of elements is quite strong and/or a large amount of elements are mutually coupled or correlated, and~this usually gives rise to unevenly distributed eigenvalues of the coupling/correlation matrix, as~will be shown by the numerical results in the subsection~below. 

\subsection{Numerical Simulations \textcolor{black}{Including~MC}}
\begin{figure}[H]
	\centering
	\includegraphics[width=0.4\columnwidth]{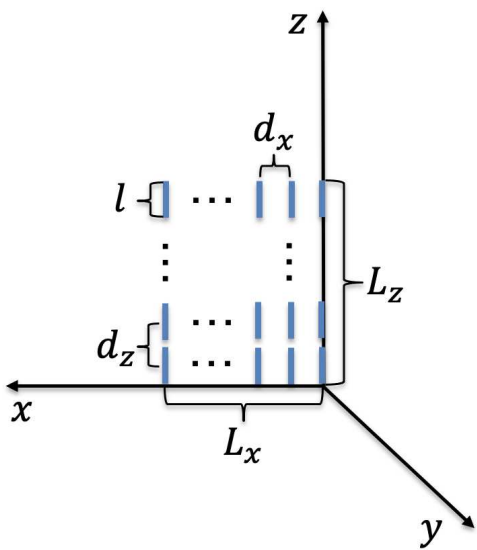}
	\caption{\textcolor{black}{Dipole topology on the $xoz$ plane with element spacing of $d_x$ and $d_z$, and~array lengths of $L_x$ and $L_z$ on the $x$- and $z$-axes, respectively. Each dipole element is of length $l=\lambda/2$.}}
	\label{fig:Dipole}
\end{figure}

Due to the complexity of the MC phenomenon, it is impossible to show generic formulas or curves for the impedance matrix or coupling matrix that apply to all types of elements or arrays. For~ease of exposition, we adopt the analytical equations for MC between identical small dipoles as an example. Although~the expressions are derived based on two dipoles, they may be applied to any pair of elements in a linear or planar array with or without a ground plane~\cite{Haupt10Book}. When the elements are small and resonant, such as dipoles, the~first-order result of MC is to alter the impedance of each of the array elements~\cite{Mailloux18Book}. Thus, the~coupling can be described in terms of mutual impedance between elements (when higher order effects can be neglected), which is also a common practice in plenty of previous research work (e.g., \cite{Malmstrom18TEC,Williams20ICCW,Gradoni21WCL}). The~mutual impedance and MC models in this paper are intended to be illustrative; for arrays composed of dipoles with other layouts or a different type of elements, it is necessary to employ more suitable mutual impedance/MC models, or~to measure these parameters in the actual array, which is deferred to future work. In~this paper, we employ the impedance matrix of half-wave dipole elements in a \textcolor{black}{parallel-in-echelon} configuration as an example, \textcolor{black}{as illustrated in Fig.~\ref{fig:Dipole}}, where $z_\text{A}\approx73.1+j42.5~\Omega$~\cite{Balanis05Book,Mailloux18Book}.  \textcolor{black}{In the simulations, $L_x=4\lambda$ and the number of half-wavelength dipoles along the $z$-axis is set to eight. The~spacing between the upper end of a dipole and the lower end of the adjacent one above it along the $z$-axis is set to $\lambda/50$, which is negligibly small compared with $\lambda$ so that $d_z\approx\lambda/2$ and $L_z\approx4\lambda$.}

\begin{figure}[H]
	\centering
	\includegraphics[width=0.7\columnwidth]{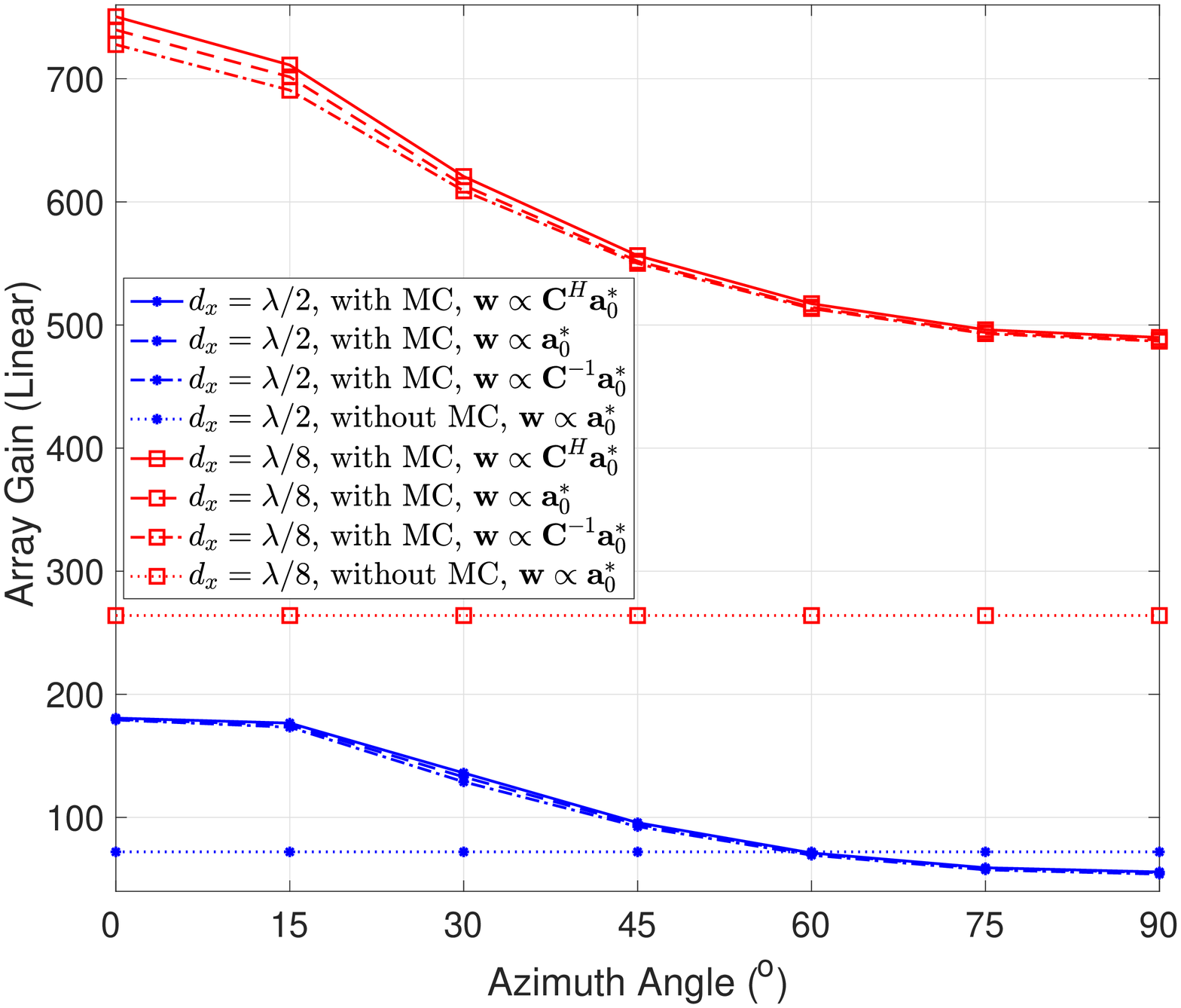}
	\caption{\textcolor{black}{Transmit (Tx) array gain as a function of the azimuth angle for both without and with mutual coupling (MC) cases. The~RIS size is $4\lambda\times4\lambda$, the~vertical element spacing is about $\lambda/2$, and~the zenith angle is set to $90^\circ$. $d_x$, $\textbf{w}$, $\textbf{C}$, and~$\textbf{a}_0$ denote the element spacing along the $x$-axis, beamforming vector, coupling matrix in~\eqref{eq:C3}, and~MC-unaware array response vector in~\eqref{eq:a2}, respectively.}}
	\label{fig:ArrayGain}
\end{figure}

As mentioned in Section~\ref{sec:SC_noMC}, one of the distinctive features of \textcolor{black}{a holographic RIS} is that its array gain may be significantly larger than a conventional array; thus, let us first investigate the array gain of holographic RISs. As~an instance, \textcolor{black}{Fig.~\ref{fig:ArrayGain} depicts the Tx array gain as a function of the azimuth angle $\phi$ (see Fig.~\ref{fig:SystemModel}) for both without and with MC cases. The~zenith angle $\theta$ (see Fig.~\ref{fig:SystemModel}) is set to $90^\circ$. $d_x$, $\textbf{w}$, $\textbf{C}$, and~$\textbf{a}_0$ denote the element spacing along the $x$-axis, beamforming vector, coupling matrix in~\eqref{eq:C3}, and~MC-unaware array response vector in~\eqref{eq:a2}, respectively. The~element spacing along the $x$-axis $d_x$ is set to $\lambda/2$ and $\lambda/8$, respectively, and~four MC plus beamforming schemes are considered for each $d_x$. Scheme 1): MC is included, and~the proposed beamforming vector $\textbf{w}$ in~\eqref{eq:w1} is adopted. Scheme 2): MC is included, and~the ordinary conjugate beamforming $\textbf{a}_0^*$ (followed by power normalization) is employed. Scheme 3): MC is included, and~the existing beamforming method in~\cite{Williams20ICCW} is utilized which maximizes the directivity of the RIS and is equivalent to $\textbf{C}^{-1}\textbf{a}_0^*$ (followed by power normalization). Scheme 4): MC is excluded, and~the optimal conjugate beamforming $\textbf{a}_0^*$ (followed by power normalization) is applied. The~following remarks can be drawn from Fig.~\ref{fig:ArrayGain}: First, comparing the cases without and with MC, the~array gain grows significantly in most cases when MC is included, even with the ordinary MC-unaware conjugate beamforming, except~for some angles with the half-wavelength spacing. For~example, the~maximum achievable array gain using the proposed beamforming method is over $2.8$ times that without MC for $d_x=\lambda/8$, and~the gap is likely to expand as the RIS becomes denser, which is promising for SNR enhancement, coverage extension, and~energy-efficient transmission for green communications. Second, when excluding MC, the~array gain ratio between $d_x=\lambda/8$ and $d_x=\lambda/2$ equals the ratio of the corresponding number of elements ($264/72\approx3.7$ herein) as expected, while this ratio reaches $8.8$ when MC is included, indicating that the array gain incorporating MC increases more rapidly with the number of elements than the without MC scenario. Third, the~proposed MC-aware beamforming approach outperforms its two counterparts including the one in~\cite{Williams20ICCW}. The~reason may lie in that the beamforming method in~\cite{Williams20ICCW} is aimed to maximize the directivity of the RIS, which represents the radiation power for a certain pointing direction against that over all pointing directions, while the proposed beamforming vector maximizes the radiation power of an array against that of a single element, i.e.,~their optimization objects are slightly different. In~general, the~performance gaps among the three beamforming schemes increase with the element density and the proximity to the end-fire direction.}

\begin{figure}[H]
	\centering
	\includegraphics[width=0.7\columnwidth]{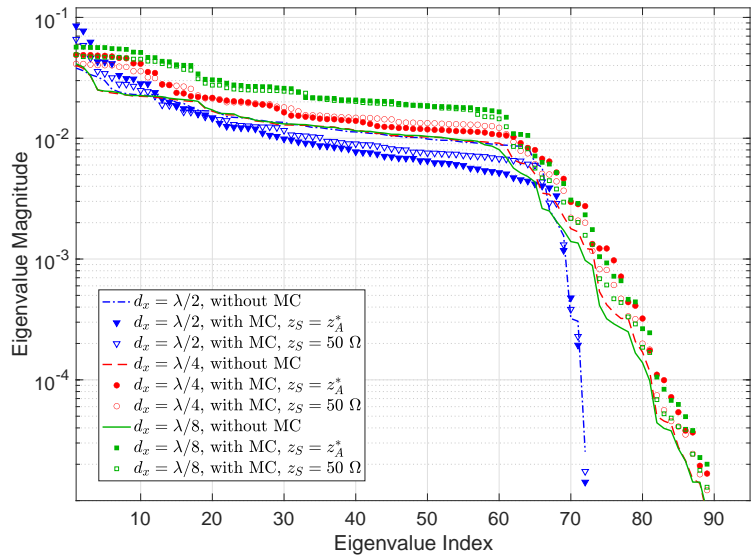}
	\caption{Eigenvalue magnitude versus eigenvalue index of the effective Tx spatial correlation matrix for horizontal element spacings of $\lambda/2$, $\lambda/4$, and~$\lambda/8$ with \textcolor{black}{a holographic RIS} size of $4\lambda\times4\lambda$, for~both without and with MC cases. $z_\text{S}$ denotes the source impedance. \textcolor{black}{Note that both the eigenvalue index and eigenvalue magnitude are dimensionless.}}
	\label{fig:EV_MC_Tx}
\end{figure}

Now, we inspect the eigenvalue behaviors of the effective spatial correlation matrix in~\eqref{eq:R3} at the Tx and Rx, respectively, for~different element intervals with a fixed \textcolor{black}{holographic RIS} size. Fig.~\ref{fig:EV_MC_Tx} depicts the eigenvalue magnitude of the effective Tx spatial correlation matrix for both without and with MC cases. Ideally, the~source impedance would be the conjugate of the element impedance, i.e.,~$z_\text{S}=z_\text{A}^*$, but~this is difficult to realize in practice. We thereby consider both perfect and imperfect impedance match by setting $z_\text{L}$ to $z_\text{A}^*$ and a common value of $50~\Omega$ to examine the influence of different impedance matching. Table~\ref{tab1} lists the associated ICSI calculated using~\eqref{eq:ICSI1}, which includes another very large source impedance of $300~\Omega$ for comparison purposes. The~following main remarks can be drawn from Fig.~\ref{fig:EV_MC_Tx} and Table~\ref{tab1}:
\begin{itemize}
	\item When excluding MC, the~most prominent inflection point of the eigenvalues in Fig.~\ref{fig:EV_MC_Tx} shifts to the left, i.e.,~the number of dominant eigenvalues decreases, as~the element spacing shrinks, which is consistent with the observations for the larger \textcolor{black}{holographic RIS} aperture in Fig.~\ref{fig:EV_12} and \textcolor{black}{is} well explained by the analysis in Section~\ref{sec:EV_noMC}.
	\item In most cases, MC increases the effective spatial correlation at Tx, as~indicated by the more rapidly-decaying eigenvalues compared with the no MC case in Fig.~\ref{fig:EV_MC_Tx} as well as the ICSI values in Table~\ref{tab1}, and~also elevates the largest eigenvalues for all element spacings studied. Furthermore, the~correlation enhancement by MC diminishes as the array becomes denser, and~MC may even have a decorrelation effect for sufficiently dense RISs.
	The reason lies in the fact that the product term $\textbf{Z}\big(\textbf{Z}+z_\text{S}\textbf{I}_N\big)^{-1}$ in~\eqref{eq:C3} approaches an identity matrix when $z_\text{S}\to0$; meanwhile, for non-zero $z_\text{S}$, it becomes a banded symmetric block-Toeplitz matrix that is sparse with non-zero entries confined to a diagonal band, and~the off-diagonal entries are significantly smaller than the diagonal ones. In addition, the~sparsity becomes more pronounced as the number of elements increases. Consequently, the~behavior of $\textbf{C}_\text{T}$ in~\eqref{eq:C3} gradually resembles that of a diagonal matrix as the element spacing dwindles, such that the effective spatial correlation becomes weaker for smaller element spacing when including MC. 
	\item Different source impedance values exert a noticeable effect on the eigenvalue structure. Specifically, the~effective spatial correlation is enhanced more substantially by perfect impedance match $z_\text{S}=z_\text{A}^*$ in contrast to $z_\text{S}=50~\Omega$, as~demonstrated by the corresponding eigenvalue trends in Fig.~\ref{fig:EV_MC_Tx} and ICSI values in Table~\ref{tab1}. This can be explained by similar reasons stated above: the properties of $\textbf{C}_\text{T}$ in~\eqref{eq:C3} are farther apart from those of a diagonal matrix with a larger $z_\text{S}$ (i.e., $73.1-j42.5~\Omega$, as~opposed to $50~\Omega$), thus higher correlation is induced. This is further verified by the ICSI for an even greater $z_\text{S}$ of $300~\Omega$ in Table~\ref{tab1}. 
	\item \textcolor{black}{In general, the~eigenvalue magnitude increases with the density of the holographic RIS, which is consistent with the array gain enhancement shown by Fig.~\ref{fig:ArrayGain}.}
\end{itemize}

\unskip

\begin{table}[H]
	\caption{Inter-element correlation/coupling strength indicator (ICSI) in~\eqref{eq:ICSI1} for Tx spatial correlation matrix without and with MC under various element~spacing.\label{tab1}}
	\newcolumntype{C}{>{\centering\arraybackslash}X}
	\begin{tabularx}{\textwidth}{CCCCC}
		\toprule
		\textbf{ICSI} \textbf{in~(\ref{eq:ICSI1})} & \textbf{Tx Spatial Correlation Matrix without MC}	& \textbf{Tx Spatial Correlation Matrix with MC, \boldmath{$z_\text{S}=z_\text{A}^*$}} & \textbf{Tx Spatial Correlation Matrix with MC, \boldmath{$z_\text{S}=50~\Omega$}} & \textbf{Tx Spatial Correlation Matrix with MC, \boldmath{$z_\text{S}=300~\Omega$}} \\
		\midrule
		$d_x=d_z=\lambda/2$	& 0.0495 & 0.0927 & 0.0752 & 0.1500\\
		$d_x=d_z=\lambda/4$	& 0.0646 & 0.0750 & 0.0665 & 0.0974\\
		$d_x=d_z=\lambda/8$	& 0.0702 & 0.0705 & 0.0671 & 0.0851\\
		\bottomrule
	\end{tabularx}
\end{table}

\begin{figure}[H]
	\centering
	\includegraphics[width=0.7\columnwidth]{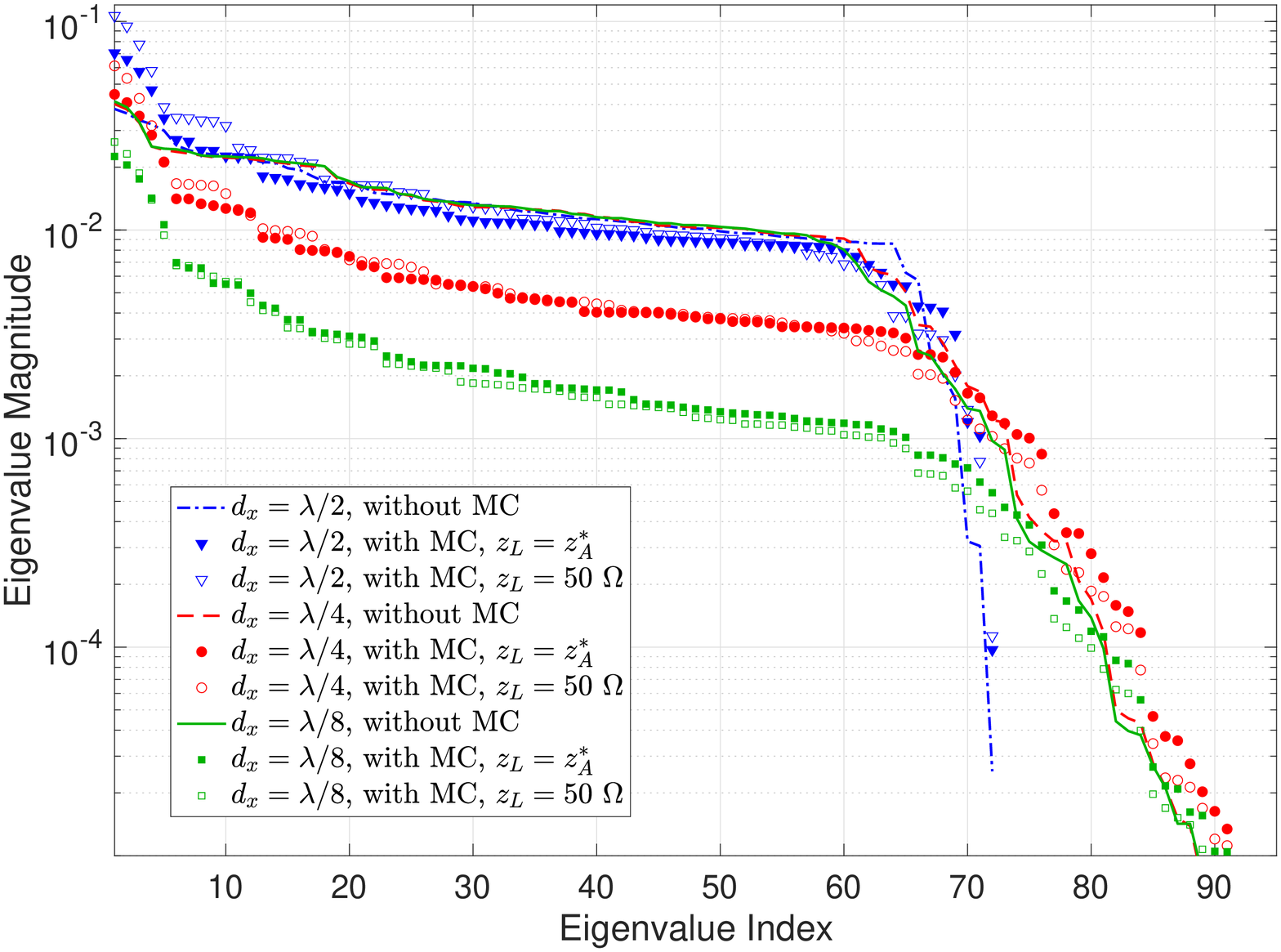}
	\caption{Eigenvalue magnitude versus eigenvalue index of the effective receive (Rx) spatial correlation matrix for horizontal element spacings of $\lambda/2$, $\lambda/4$, and~$\lambda/8$ with \textcolor{black}{a holographic RIS} size of $4\lambda\times4\lambda$, for~both without and with MC cases. $z_\text{L}$ denotes the load impedance. \textcolor{black}{Note that both the eigenvalue index and eigenvalue magnitude are dimensionless.}}
	\label{fig:EV_MC_Rx}
\end{figure}

The eigenvalue magnitude of the effective spatial correlation matrix at the Rx side is depicted in Fig.~\ref{fig:EV_MC_Rx} under the same simulation settings as in Fig.~\ref{fig:EV_MC_Tx}, and~the corresponding ICSI values are provided in Table~\ref{tab2}. Some interesting phenomena are observed and described~below.
\unskip
\begin{table}[H]
	\caption{ICSI {in~\eqref{eq:ICSI1}} for Rx spatial correlation matrix without and with MC under various element~spacing.\label{tab2}}
	\newcolumntype{C}{>{\centering\arraybackslash}X}
	\begin{tabularx}{\textwidth}{CCCCC}
		\toprule
		\textbf{ICSI} \textbf{in~{(\ref{eq:ICSI1})}} & \textbf{Rx Spatial Correlation Matrix without MC}	& \textbf{Rx Spatial Correlation Matrix with MC, \boldmath{$z_\text{L}=z_\text{A}^*$}} & \textbf{Rx Spatial Correlation Matrix with MC, \boldmath{$z_\text{L}=50~\Omega$}} & \textbf{Rx Spatial Correlation Matrix with MC, \boldmath{$z_\text{L}=300~\Omega$}} \\
		\midrule
		$d_x=\lambda/2$		& 0.0495 & 0.0682 & 0.0787 & 0.0550\\
		$d_x=\lambda/4$		& 0.0646 & 0.0867 & 0.1001 & 0.0698\\
		$d_x=\lambda/8$		& 0.0702 & 0.1045 & 0.1213 & 0.0828\\
		\bottomrule
	\end{tabularx}
\end{table}

\begin{itemize}
	\item MC at the Rx reduces the magnitude of eigenvalues in most cases (except for the first few largest eigenvalues). Compared with the Tx coupling matrix $\textbf{C}_\text{T}$ in~\eqref{eq:C3}, the~Rx coupling matrix $\textbf{C}_\text{R}$ in~\eqref{eq:C5} is mainly different by a term $z_\text{A}\textbf{Z}^{-1}$, which causes the reduction of the eigenvalue magnitude after multiplying with the original spatial correlation matrix. 
	\item As seen in Table~\ref{tab2}, MC increases the effective spatial correlation at the Rx, which is similar to the Tx side. However, contrary to the observation for the Tx side, MC with a larger load impedance tends to have less correlation enhancement effect at Rx in general, as~shown by Fig.~\ref{fig:EV_MC_Rx} and Table~\ref{tab2}, since a larger load impedance renders the term $z_\text{L}\textbf{I}_N$ more dominant in $\left(\textbf{Z}+z_\text{L}\textbf{I}_N\right)^{-1}$ in~\eqref{eq:C5} so that the coupling matrix is more like a diagonal matrix with smaller off-diagonal entries and hence weaker correlation. 
	\item Rx MC increases the effective spatial DoF for element spacings less than half a wavelength, as~implied by the most prominent inflection point of the eigenvalues which shifts to the right when considering MC. This is beneficial for spatial diversity and multistreaming. 
\end{itemize}

\textcolor{black}{To analyze and compare the MC effects of different element types, we also look at the element with an isotropic radiation pattern which is commonly assumed in the existing work. The~real part of the impedance matrix of isotropic elements is~\cite{Ivrlac10TCS}}
\begin{equation}\label{eq:Z2}\color{black}
\begin{split}
[\mathfrak{R}\{\textbf{Z}_\text{iso}\}]_{n_1,n_2}=r_\text{iso}\text{sinc}\bigg(\frac{2||\textbf{d}_{n_1}-\textbf{d}_{n_2}||_2}{\lambda}\bigg),~n_1,n_2=1,\ldots,N
\end{split}
\end{equation}

\begin{figure}[H]
	\centering
	\includegraphics[width=0.7\columnwidth]{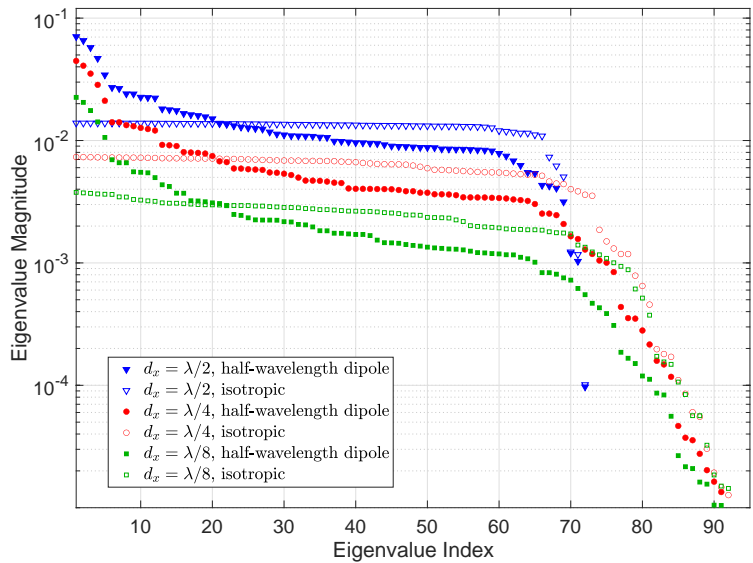}
	\caption{\textcolor{black}{Eigenvalue magnitude versus eigenvalue index of the effective Rx spatial correlation matrix for half-wavelength dipoles and isotropic elements. The~element spacing along the $x$-axis is set to $\lambda/2$, $\lambda/4$, and~$\lambda/8$, respectively, with a holographic RIS size of $4\lambda\times4\lambda$, for~both without and with MC cases. The~load impedance $z_\text{L}=z_\text{A}^*$. Note that both the eigenvalue index and eigenvalue magnitude are dimensionless.}}
	\label{fig:EV_MC_Rx_iso}
\end{figure}

\noindent \textcolor{black}{where $r_\text{iso}$ denotes the radiation resistance of an isotropic element, and the definitions of $\text{sinc}(x)$, $\textbf{d}_{n_1}$, and~$\textbf{d}_{n_2}$ are aligned with those in~\eqref{eq:R5}. By applying appropriate matching techniques, the imaginary part of the impedance matrix can be removed while the real part remains~\cite{Ivrlac10TCS}, thus we only need to consider the real part. The~Rx coupling matrix for isotropic elements is obtained by plugging~\eqref{eq:Z2} into~\eqref{eq:C5} and setting $z_\text{L}=r_\text{iso}$. Fig.~\ref{fig:EV_MC_Rx_iso} displays the eigenvalue magnitude versus eigenvalue index of the effective Rx spatial correlation matrix for half-wavelength dipoles and isotropic elements, which shows that the effective spatial correlation for isotropic elements is obviously lower than that for half-wavelength dipoles, as~manifest from the more evenly distributed eigenvalue magnitude for isotropic elements. This is because the MC is zero whenever the spacing between two isotropic elements is multiples of half a wavelength, as~indicated by~\eqref{eq:Z2}, while the MC is all non-zero for the dipoles herein, hence the overall MC is lower for isotropic elements.}
\unskip

\section{Conclusions}
\label{sec:Conclusions}
In this paper, we have investigated the array response and small-scale spatial correlation for \textcolor{black}{holographic RISs} excluding and including MC. In-depth analysis is conducted on the asymptotic eigenvalue distribution and spatial DoF for the spatial correlation matrix under isotropic scattering, by~linking the eigenvalues to the power spectrum of the spatial correlation function based on the BTTB matrix theory. It is demonstrated that the specious more dominant eigenvalues for fewer elements in \textcolor{black}{holographic RISs} are due to the spectrum aliasing in the wavenumber domain which enhances the near-field evanescent waves, thus the far-field spatial DoF actually does not increase. Furthermore, \textcolor{black}{an MC-aware beamforming scheme is proposed which is aimed to maximize the array gain, and~is shown to outperform existing methods}. In~addition, it is found that MC exerts discrepant effects on Tx and Rx modes: For the Tx, MC increases the eigenvalue magnitude and effective spatial correlation in most cases, while Rx MC often reduces the eigenvalue magnitude while increasing the spatial DoF, especially for dense \textcolor{black}{RISs}. The~array gain and channel eigenvalue enhancement by Tx MC is beneficial to SNR elevation, coverage extension, and~energy saving, and~the growing spatial DoF caused by Rx MC indicates that efficient spatial multiplexing is possible even with compact arrays. 

\bibliographystyle{IEEEtran}
\bibliography{myref}

\end{document}